\documentclass[prd,twocolumn,amsmath,nobibnotes,nofootinbib,superscriptaddress,preprintnumbers]{revtex4-1}

\usepackage{amssymb}
\usepackage{amsmath}
\usepackage{graphicx}
\usepackage{subfigure}
\usepackage{booktabs}
\usepackage{ifpdf}
\usepackage{color} 
\ifpdf
  \usepackage[pdftex]{hyperref}
\else
  \usepackage[ps2pdf,colorlinks=true]{hyperref}
\fi

\newif\ifarxiv
\arxivtrue

\newcommand{\eqn}[1]{(#1)}

\newcommand{\tbl}[1]{Table~#1}

\newcommand{\fig}[1]{Fig.~#1}

\newcommand{\sectn}[1]{Sec.~#1}

\newcommand{\eg}{\mbox{\it e.g.}}
\newcommand{\ie}{\mbox{\it i.e.}}
\newcommand{\etc}{\mbox{\it etc.}}

\newcommand{\kelvin}{{K}}
\newcommand{\degrees}{\ensuremath{{^\circ}}}

\newcommand{\cmb}{{CMB}}
\newcommand{\cmbtext}{{cosmic microwave background}}

\newcommand{\wmap}{{WMAP}}
\newcommand{\wmaptext}{{Wilkinson Microwave Anisotropy Probe}}

\newcommand{\healpix}{{\tt HEALPix}}

\newcommand{\stwo}{{\tt S2}}
\newcommand{\stwofil}{{\tt S2FIL}}

\newcommand{\comb}{{\tt COMB}}

\newcommand{\fwhm}{{FWHM}}

\newcommand{\snr}{{\rm SNR}}

\newcommand{\mf}{{MF}}

\newcommand{\spcend}{\ensuremath{\:}}

\newcommand{\cconj}{\ensuremath{\ast}}

\newcommand{\sphere}{\ensuremath{{\mathrm{S}^2}}}
\newcommand{\sothree}{\ensuremath{{\mathrm{SO}(3)}}}

\newcommand{\opnexpv}{\ensuremath{\langle}}
\newcommand{\clsexpv}{\ensuremath{\rangle}}

\newcommand{\dx}{\ensuremath{\mathrm{\,d}}}
\newcommand{\dmu}[1]{\ensuremath{\dx \Omega(#1)}}

\newcommand{\innerp}[2]{\ensuremath{\langle {#1},\: {#2} \rangle}}

\newcommand{\saa}{\ensuremath{\theta}}
\newcommand{\sab}{\ensuremath{\varphi}}
\newcommand{\sas}{\ensuremath{\saa, \sab}}
\newcommand{\eul}{\ensuremath{\mathbf{\rho}}}

\newcommand{\el}{\ensuremath{\ell}}
\newcommand{\m}{\ensuremath{m}}

\newcommand{\elmax}{\ensuremath{{L}}}
\newcommand{\p}{\ensuremath{^\prime}}

\newcommand{\scale}{\ensuremath{R}}

\newcommand{\kron}[2]{\ensuremath{\delta_{{#1}{#2}}}}

\newcommand{\shfarg}[3]{\ensuremath{Y_{#1#2}({#3})}}

\newcommand{\shf}[2]{\ensuremath{Y_{#1#2}}}

\newcommand{\shc}[3]{\ensuremath{{#1}_{{#2}{#3}}}}

\newcommand{\rot}{\ensuremath{\mathcal{R}}}

\newcommand{\sky}{\ensuremath{f}}

\newcommand{\sumlm}{\ensuremath{\sum_{\el=0}^{\infty} \sum_{\m=-\el}^\el}}

\newcommand{\summ}{\ensuremath{\sum_{\m=-\el}^\el}}

\newcommand{\sumlmb}{\ensuremath{\sum_{\el \m}}}

\newcommand{\order}{\ensuremath{\mathcal{O}}}

\newcommand{\alm}{\ensuremath{a}}
\newcommand{\almi}{\ensuremath{\shc{\alm}{\el}{\m}}}

\newcommand{\fil}{\ensuremath{\varphi}}
\newcommand{\filcoeff}{\ensuremath{w}}

\newcommand{\filvara}{\ensuremath{a}}

\newcommand{\scalepnorm}{\ensuremath{{\scale\pnormsep\pnorm}}}

\newcommand{\detect}{\ensuremath{\Gamma}}

\newcommand{\noise}{\ensuremath{n}}

\newcommand{\amp}{\ensuremath{A}}
\newcommand{\tmpl}{\ensuremath{\tau}}
\newcommand{\noisecl}{\ensuremath{C}}

\newcommand{\thetacrit}{\ensuremath{\saa_{\rm crit}}}
\newcommand{\zo}{\ensuremath{z_{0}}}
\newcommand{\zcrit}{\ensuremath{z_{\rm crit}}}
\newcommand{\saao}{\ensuremath{\saa_{0}}}
\newcommand{\sabo}{\ensuremath{\sab_{0}}}
\newcommand{\saso}{\ensuremath{\saao, \sabo}}

\renewcommand{\sab}{\ensuremath{\phi}}
\renewcommand{\sky}{\ensuremath{\Delta T}}
\renewcommand{\fil}{\ensuremath{\Psi}}
\renewcommand{\filcoeff}{\ensuremath{F}}
\renewcommand{\filvara}{\ensuremath{\alpha}}
\renewcommand{\elmax}{\ensuremath{\el_{\rm max}}}
\renewcommand{\scalepnorm}{\ensuremath{\scale}}
\renewcommand{\opnexpv}{\ensuremath{\mathbb{E} [ }}
\renewcommand{\clsexpv}{\ensuremath{]}}
\newcommand{\refn}[1]{Ref.~#1}

\newcommand{\refns}[1]{Refs.~#1}

\renewcommand{\eqn}[1]{Eq.~(#1)}

\begin{document}

\title{Optimal filters for detecting cosmic bubble collisions}
\date{\today}

\author{J.~D.~McEwen}
\email{jason.mcewen@ucl.ac.uk}
\affiliation{Department of Physics and Astronomy, University College London, London WC1E 6BT, U.K.}
\author{S.~M.~Feeney}
\email{stephen.feeney.09@ucl.ac.uk}
\affiliation{Department of Physics and Astronomy, University College London, London WC1E 6BT, U.K.}
\author{M.~C.~Johnson}
\email{mjohnson@perimeterinstitute.ca}
\affiliation{Perimeter Institute for Theoretical Physics, Waterloo, Ontario N2L 2Y5, Canada} 
\author{H.~V.~Peiris}
\email{h.peiris@ucl.ac.uk}
\affiliation{Department of Physics and Astronomy, University College London, London WC1E 6BT, U.K.}

\begin{abstract}
A number of well-motivated extensions of the $\Lambda$CDM concordance cosmological model postulate the existence of a population of sources embedded in the \cmbtext\ (\cmb). One such example is the signature of cosmic bubble collisions which arise in models of eternal inflation. The most unambiguous way to test these scenarios is to evaluate the full posterior probability distribution of the global parameters defining the theory; however, a direct evaluation is computationally impractical on large datasets, such as those obtained by the \wmaptext\ (\wmap) and Planck. A method to approximate the full posterior has been developed recently, which requires as an input a set of candidate sources which are most likely to give the largest contribution to the likelihood. In this article, we present an improved algorithm for detecting candidate sources using optimal filters, and apply it to detect candidate bubble collision signatures in \wmap\ \mbox{7-year} observations. We show both theoretically and through simulations that this algorithm provides an enhancement in sensitivity over previous methods by a factor of approximately two. Moreover, no other filter-based approach can provide a superior enhancement of these signatures. Applying our algorithm to \wmap\ \mbox{7-year} observations, we detect eight new candidate bubble collision signatures for follow-up analysis.
\end{abstract}

\preprint{}

\maketitle

\section{Introduction}

Precision observations of the \cmbtext\ (\cmb) provide the most accurate picture of the early universe that is available currently. The standard $\Lambda$CDM concordance cosmological model -- which states that we live in a universe composed mostly of dark energy and dark matter, whose structure was seeded by adiabatic and very nearly Gaussian and scale-invariant density perturbations -- describes the statistics of temperature fluctuations in the \cmb\ extremely well~\cite{komatsu:2010,larson:2011}. However, there are many theoretically well-motivated extensions of $\Lambda$CDM that predict detectable secondary signals in the \cmb. 

One example, which has been the subject of a number of recent studies~\cite{Garriga:2006hw,Aguirre:2007an,Aguirre:2007wm,Aguirre:2008wy,Chang:2007eq,Chang:2008gj,Czech:2010rg,Dahlen:2008rd,Freivogel:2009it,Larjo:2009mt,Kleban:2011yc,Gobbetti:2012yq}, is the signature of cosmic bubble collisions which arise in models of eternal inflation (see \refn{\cite{Aguirre:2009ug}} for a review).  In the model of eternal inflation, our observable universe is contained inside one member of an ensemble of bubbles. Collisions between bubbles disturb the homogeneity and isotropy of the very early universe, leaving possibly detectable imprints on the \cmb. In the limit where the number of detectable collisions on the \cmb\ sky is relatively small, the signature is a set of azimuthally-symmetric modulations of the temperature~\cite{Garriga:2006hw,Aguirre:2007an}, varying as the cosine of the angular distance from the collision centre~\cite{Chang:2008gj}, with a size-distribution peaking at half-sky scales~\cite{Freivogel:2009it}. Other examples of secondary signals arise in theories with topological defects such as cosmic strings (see \eg~\refn{\cite{Vilenkin:1986hg}} for a review) or textures~\cite{turok:1990}; a less exotic example is the signature of clusters of galaxies produced by the Sunyaev-Zel'dovich (SZ) effect~\cite{sz:1980}. 

In each of these examples, a population of sources is hypothesized to exist on top of the background \cmb, the members of which have properties drawn from a calculable probability distribution. The most unambiguous way to test these scenarios is to utilize the most general predictions for the population of sources on the full-sky, and determine the posterior probability distribution over the global parameters defining the theory (such as the total number of features expected, their intrinsic amplitude, \etc). The enormous size of modern \cmb\ datasets, such as those obtained by the \wmaptext\ \cite{bennett:2003c} (\wmap) and those currently being obtained by the Planck satellite~\cite{tauber:2010}, provide a unique challenge for such an analysis. Indeed, a direct pixel-based evaluation of the posterior at full resolution is computationally intractable. 

Recently, however, \refns{\cite{feeney:2011a,feeney:2011b}} outlined a method for approximating the full posterior describing source populations in the context of the bubble collision hypothesis. The method is generalized to the detection of other sources easily. This approach requires pre-processing of the data to recover a set of candidate sources which are most likely to give the largest contribution to the likelihood.  The pre-processing stage of this method is thus crucial to its overall effectiveness.  Candidate source detection aims to minimize the number of false detections while remaining sensitive to a weak signal; a manageable number of false detections is thus tolerated, as the subsequent Bayesian processing step will discriminate these from true signals.  To detect candidate bubble collision signatures, \refns{\cite{feeney:2011a,feeney:2011b}} employ a suite of needlet transforms~\cite{marinucci:2008,scodeller:2010}.  Needlets are a form of azimuthally-symmetric wavelet\footnote{Note that Mexican needlets \cite{scodeller:2010} are not formally wavelets since exact synthesis is not possible, even in theory.} defined on the sphere, that render the location and scale of candidate features simultaneously accessible\footnote{Needlets are in fact the azimuthally-symmetric restriction of exact steerable wavelets defined on the sphere \cite{wiaux:2007:sdw}, which render the orientation of directional features also accessible.}. While the effectiveness of needlets for detecting candidate features has been demonstrated already~\cite{feeney:2011a,feeney:2011b}, needlets are generic and are not adapted to the signal of interest; consequently, they are not optimal. A better approach is to enhance the effectiveness of candidate detection by exploiting knowledge of the source signature.

Optimal filters have found widespread application in many branches of physics and signal processing for the detection of compact objects embedded in a stochastic background.  In the context of astrophysics, the matched filter has been applied to detect point sources and SZ emission in \cmb\ observations~\cite{tegmark:1998,haehnelt:1995}.  Alternative optimal filters, such as the scale-adaptive filter, have also been derived~\cite{sanz:2001,herranz:2002}~and applied to \cmb\ observations~\cite{barreiro:2003}.  In all of these cases, optimal filters are applied to small patches of the sky, where a flat tangent plane approximation of the celestial sphere in the region of interest is made.  To analyze full-sky \cmb\ observations these techniques must be extended from Euclidean space to a spherical manifold.  Optimal filter theory has been extended to the sphere by \refn{\cite{schaefer:2004}} (and applied to detect SZ emission~\cite{schaefer:2006}) for the case of azimuthally-symmetric source signatures and by \refn{\cite{mcewen:2006:filters}} for the general directional setting.

In this article we develop an alternative candidate source detection algorithm using optimal filters. We focus on the problem of detecting the signatures of bubble collisions in observations of the \cmb, but our approach generalizes to other sources and backgrounds trivially. Since the angular scale of a typical bubble collision is expected to be large~\cite{Freivogel:2009it,Aguirre:2007an,Aguirre:2009ug}, tangent plane approximations are not valid, and we instead consider optimal filters defined on the sphere~\cite{schaefer:2004,mcewen:2006:filters}.  We describe and evaluate our new candidate source detection algorithm in \sectn{\ref{sec:filters}} and show it to be superior to the needlet approach considered previously~\cite{feeney:2011a,feeney:2011b}.  Finally, we apply our algorithm to \wmap\ observations in \sectn{\ref{sec:wmap}}, resulting in the detection of a number of new candidate bubble collision signatures in the \wmap\ \mbox{7-year} data. Concluding remarks are made in \sectn{\ref{sec:conclusions}}.

\section{Optimal detection of candidate bubble collisions}
\label{sec:filters}

Filter based approaches to enhance a signal in a background process are common due to their effectiveness and efficiency.  Indeed, a wavelet transform, such as needlets, is merely a filtering operation with a carefully constructed set of filter kernels (to allow the exact reconstruction of the original signal).  In this section we consider filters that provide the maximal enhancement of the source signature in a given stochastic background.  The filters are optimal in the sense that no other filter can yield a greater enhancement in the signal-to-noise ratio (\snr) of the filtered field.  Our optimal-filter-based method is general: in this work, we focus on its application to the problem of detecting signatures of bubble collisions.  Firstly, we define the signatures of the bubble collision remnants that we search for.  We then construct and evaluate optimal filters for detecting candidate bubble collision signatures when the size of the signature is known, before describing an algorithm for detecting multiple candidate bubble collision signatures of unknown and differing sizes.

\subsection{Bubble collision signatures}

\begin{figure*}
\centering
\mbox{
\subfigure[Bubble collision signature radial profile]{\includegraphics[height=45mm]
  {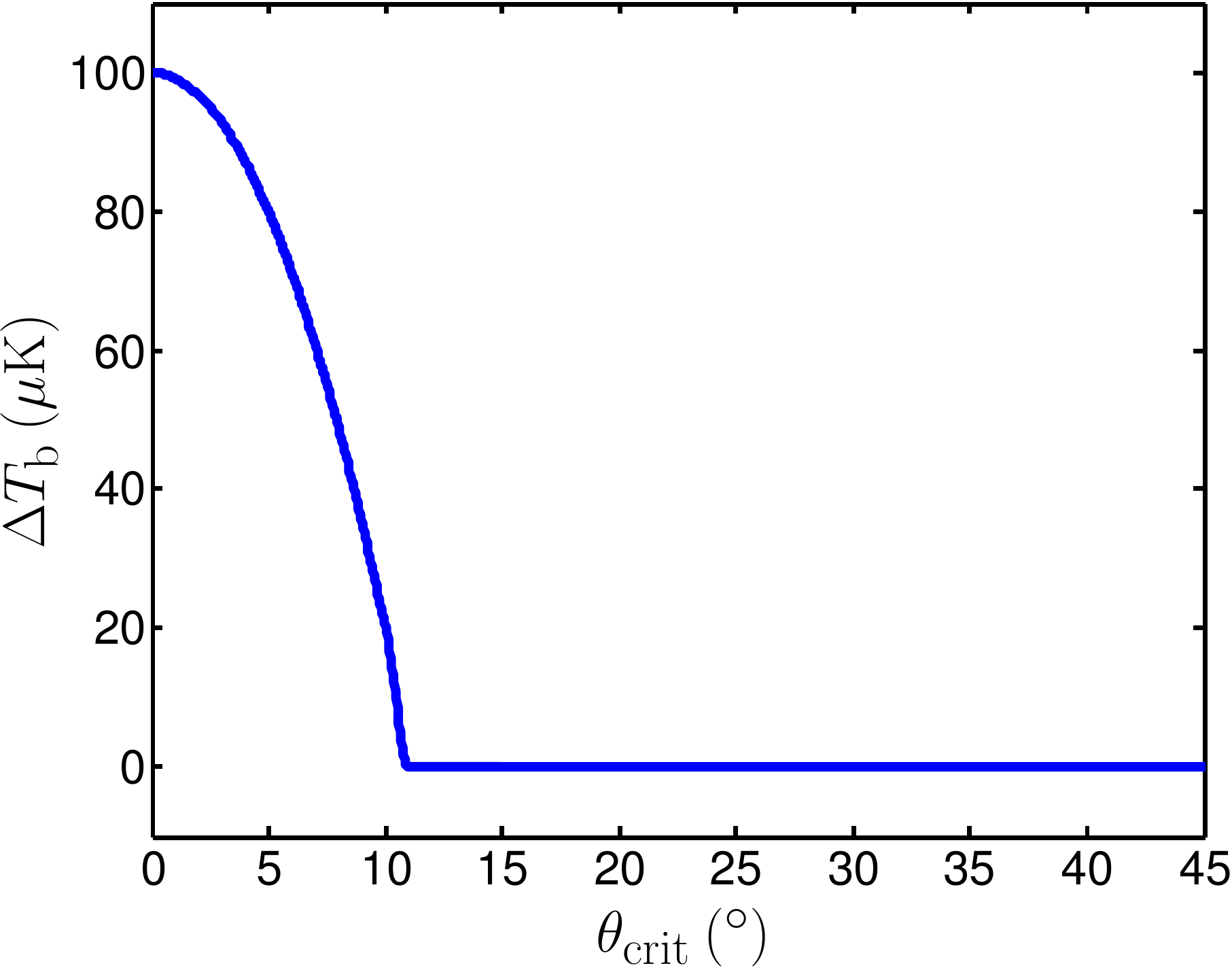}}
\quad\quad
\subfigure[Bubble collision signature on the sphere]{
  \includegraphics[viewport=0 0 220 220, clip=, height=45mm]
  {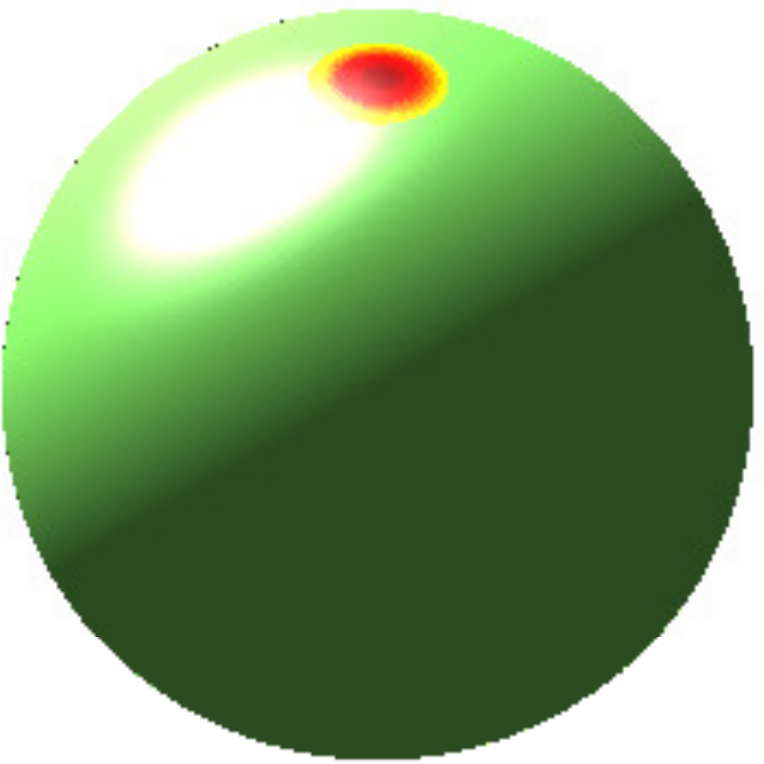}}
\quad\quad
\subfigure[Power spectrum of bubble collision\newline signature and \cmb]{\includegraphics[height=45mm]
  {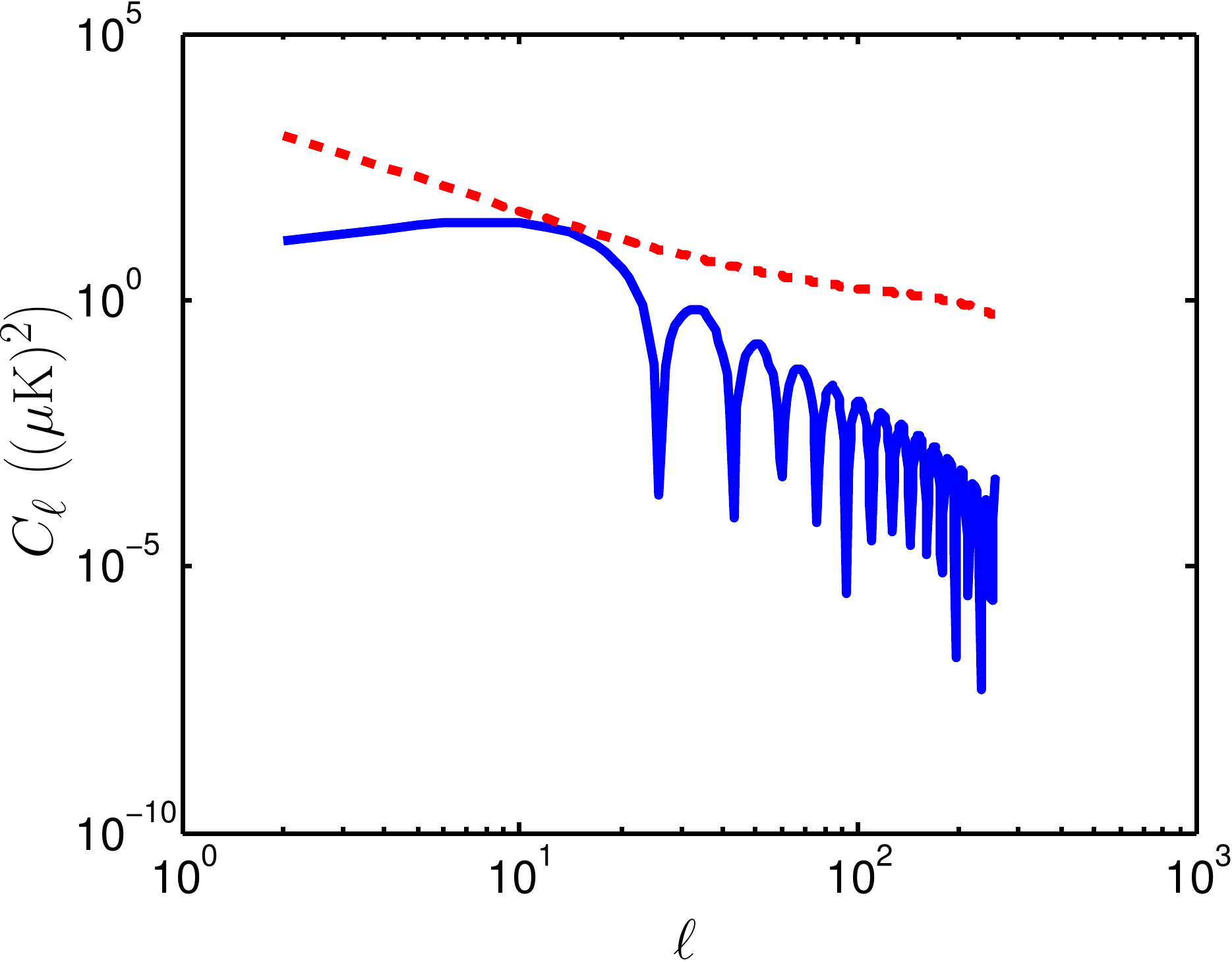}}
}
\caption{Panels~(a) and~(b) show the radial profile and 3D surface plot (lit from top-left, with specular highlight), respectively, of a bubble collision signature with parameters $\{\zo,\thetacrit,\saso \}=\{ 100\ \mu{\rm \kelvin, 10\degrees, 0\degrees, 0\degrees}\}$ (throughout we consider \mbox{$\zcrit\sim0\ \mu$\kelvin).} In panel~(c) the power spectrum of the bubble collision signature (solid blue curve) is compared with the best-fit WMAP7$+$BAO$+$H0 \cmb\ power spectrum (red dashed curve). Matched filters for azimuthally symmetric templates promote harmonic modes where the source template power spectrum is large and suppress modes where the \cmb\ power spectrum is large.}
\label{fig:template}
\end{figure*}

Bubble collisions induce a modulative and additive contribution to the temperature fluctuations of the \cmb~\cite{Chang:2008gj}, however the modulative component is second order and may be safely ignored. The additive contribution induced in the \cmb\ by a bubble collision is given by the azimuthally-symmetric profile
\begin{equation*}
\sky_{\rm b}(\sas) =  \left[ c_0 + c_1 \cos ( \saa) \right] s(\saa; \thetacrit) \spcend ,
\end{equation*}
when centered on the North pole, where $(\sas)\in\sphere$ denote the spherical coordinates of the unit sphere \sphere, with colatitude $\saa \in [0,\pi]$ and longitude $\sab \in [0,2\pi)$, and $c_0$ and $c_1$ are free parameters (not to be confused with the power spectrum monopole and dipole).  A typical bubble collision signature is illustrated in \fig{\ref{fig:template}}.  Following the parameterization of \refns{\cite{feeney:2011a,feeney:2011b}}, we describe the bubble collision signature by its amplitude at its centre and at its causal boundary, given by $\zo = c_0 + c_1$ and $\zcrit = c_0 + c_1 \cos(\thetacrit)$ respectively, and by its size $\thetacrit$.  We replace the discontinuous Heaviside step function of the bubble collision profile with a ``Schwartz'' step function $s(\saa; \thetacrit)$ that is infinitely differentiable but nevertheless exhibits a smooth but rapid transition to zero about $\thetacrit$. As theoretical work suggests that the temperature discontinuity parameter should be negligible~\cite{Gobbetti:2012yq,Kleban:2011yc} (an observation that is supported by the candidate bubble collision signatures detected previously~\cite{feeney:2011a,feeney:2011b}), we restrict our attention to $\zcrit \sim 0\ \mu \kelvin$.  Bubble collision signatures may occur at any position on the sky $(\saso)$ and at a range of sizes $\thetacrit$ and amplitudes $\zo$.  We denote by $\sky_{i}$ the temperature contribution induced by a candidate bubble collision $i$ with parameters $\{\zo, \thetacrit, \saso\}$.

\subsection{Optimal bubble collision filters}

We define optimal filters to enhance the contributions of compact sources embedded in a stochastic background, focusing on the case of locating candidate bubble collision signatures in the \cmb.  Firstly, we discuss filtering on the sphere in general, before defining the optimal matched filter.  We compute the matched filter for detecting bubble collision signatures and compare the \snr\ for the matched filter to alternatives, such as needlets and the unfiltered field itself.

\subsubsection{Filtering}

Filtering on the sphere is the natural analogue of the filtering operation in Euclidean space and is defined by the projection of a function, such as the \cmb\ temperature fluctuations $\sky$, onto rotated filter kernels.  Consequently, filtering on the sphere is defined through the spherical convolution
\begin{align}
\label{eqn:filtering}
\displaystyle
\filcoeff_\scale(\eul) &=
\innerp{\sky}{\rot(\eul)\fil_{\scalepnorm}} \\
&=
\int_{\sphere}
\dmu{\saa\p,\sab\p} \:
\sky(\saa\p,\sab\p) \:
[\rot(\eul) \fil_\scalepnorm]^\cconj(\saa\p,\sab\p)
\spcend , \nonumber
\end{align}
where $\fil_\scale$ is the filter kernel at scale \scale, $\rot$ is the rotation operator describing a rotation by the Euler angles $\eul\in\sothree$, $\innerp{\cdot}{\cdot}$ denotes the inner product on the sphere, \mbox{${}^\cconj$ denotes} complex conjugation and $\dmu{\saa,\sab} = \sin\saa \dx\saa\dx\sab$ is the usual rotation-invariant measure on the sphere.  The filtering operation given by \eqn{\ref{eqn:filtering}} is general in the sense that directional filter kernels are considered.  Since we are concerned with bubble collision signatures, which are azimuthally-symmetric, we henceforth restrict our attention to azimuthally-symmetric filter kernels such that $\fil_\scale(\sas) = \fil_\scale(\saa)$.  In this case, the filter kernel is invariant under rotations about its own axis of symmetry and the set of distinct rotations is restricted from the rotation group \sothree\ to the sphere \sphere, \ie\ $\eul=(\sas)\in\sphere$.  

Just like in the Euclidean setting, filtering on the sphere can be computed more efficiently in harmonic space than through an evaluation of \eqn{\ref{eqn:filtering}} by direct quadrature.  The \cmb\ temperature fluctuations may be represented by their expansion in the basis of spherical harmonics \shf{\el}{\m}, given by
\begin{equation}
\label{eqn:harmonic_expansion}
\sky(\sas) = \sumlm \almi \shfarg{\el}{\m}{\sas}
\spcend ,
\end{equation}
where the harmonic coefficients are given by the usual projection onto the basis functions: $\almi = \innerp{\sky}{\shf{\el}{\m}}$.  In practice, we consider a maximum band-limit $\elmax$, such that the summation over $\el$ in \eqn{\ref{eqn:harmonic_expansion}} may be truncated to $\elmax$.  Similarly, the filter kernel may be decomposed into its spherical harmonic expansion, with coefficients given by $\shc{(\fil_\scalepnorm)}{\el}{\m} = \innerp{\fil_\scale}{\shf{\el}{\m}}$.  For an azimuthally-symmetric kernel the filter coefficients are non-zero for harmonic indices $\m=0$ only, \ie\ $\shc{(\fil_\scalepnorm)}{\el}{\m} = \kron{\m}{0} \shc{(\fil_\scalepnorm)}{\el}{0}$, where $\kron{i}{j}$ is the Kronecker delta symbol.  In this setting, the harmonic coefficients of the filtered field are given by
\begin{equation}
\label{eqn:filtering_harmonic}
\shc{(\filcoeff_\scale)}{\el}{\m} = \sqrt{\frac{4\pi}{2\el+1}} \: \almi \: \shc{(\fil_\scalepnorm)}{\el}{0}^\cconj
\spcend .
\end{equation}
Fast spherical harmonic transforms (\eg\  \refns{\cite{driscoll:1994,gorski:2005,doroshkevich:2005,mcewen:fssht}}) may then be employed to reduce the complexity of filtering with an azimuthally-symmetric kernel from $\order(\elmax{}^4)$ to $\order(\elmax{}^3)$.\footnote{Filtering with directional filter kernels can also be performed more efficiently in harmonic space than in real space \cite{risbo:1996,wandelt:2001,mcewen:2006:fcswt}.}

The purpose of filtering the observed signal on the sphere is to enhance source signatures relative to the stochastic background; we thus require a quantitive measure of the effectiveness of filtering.  We define the \snr\ of the filtered field for scale \scale\ by the ratio of its mean to its dispersion in the presence of a source located at $(\saso)$:
\begin{equation}
\label{eqn:snr}
\detect_\scale =
\frac{\mu_\scale(\saso)}
{\sigma_{\scale}(\saso)}
\spcend ,
\end{equation}
where the mean and variance of the filtered field are defined, respectively, by 
\begin{equation*}
\mu_\scale(\sas) = 
\opnexpv
 \filcoeff_\scale(\sas)
\clsexpv
\end{equation*}
and
\begin{equation*}
\sigma_\scale^2(\sas) = 
\opnexpv
| \filcoeff_\scale(\sas) |^2
\clsexpv
- \mu_\scale^2(\sas)
\spcend .
\end{equation*}

\subsubsection{Optimal filters}

The observed \cmb\ temperature fluctuations $\sky$ are assumed to be comprised of a number of compact sources $\sky_{i}$, such as bubble collision signatures, embedded in a stochastic background noise process $\noise$:
\begin{equation*}
\sky(\sas) = \sum_i \sky_{i}(\sas) + \noise(\sas)
\spcend .
\end{equation*}
We decompose the sources into their amplitude $\amp_i$ and normalized template profile $\tmpl_i$ by $\sky_i(\sas) = \amp_i \: \tmpl_i(\sas)$; for the case of bubble collision signatures we make the association $\amp = \zo$.  The stochastic noise process $\noise$ is assumed to be zero-mean, isotropic and homogeneous and is defined by its power spectrum:
\begin{equation*}
\opnexpv
\shc{\noise}{\el}{\m}
\shc{\noise}{\el\p}{\m\p}^\cconj
\clsexpv
=
\noisecl_\el \:
\kron{\el}{\el\p} \:
\kron{\m}{\m\p}
\spcend ,
\end{equation*}
where $\shc{\noise}{\el}{\m} = \innerp{\noise}{\shf{\el}{\m}}$.  The source population is the signal of interest, hence the noise is comprised of primary and secondary \cmb\ anisotropies.

We filter the observed \cmb\ temperature fluctuations \sky\ with the aim of enhancing the source contributions $\sky_i$ relative to the background noise $\noise$.  The matched filter $\fil_\scalepnorm^{\rm MF}$ is defined to maximize the \snr\ of the filtered field given by \eqn{\ref{eqn:snr}}, while ensuring that the amplitude of the filtered field at the source position gives an unbiased estimator of the source amplitude.  Thus, the matched filter defined on the sphere is recovered by solving the constrained optimization problem:
\begin{equation*}
  \min_{{\rm w.r.t.}\: \fil_\scale } \:
  {\sigma_\scale^2(\saso)} 
  \:\:
  \mbox{such that}
  \:\:
  \mu_\scale(\saso) = \amp
  \spcend .
\end{equation*}
The resulting matched filter is given by \cite{mcewen:2006:filters}
\begin{equation}
\label{eqn:mf}
  \shc{(\fil_\scale^{\rm MF})}{\el}{m} =
  \frac{\shc{\tmpl}{\el}{m}}
  {
    \filvara \:
    \noisecl_\el
  }
  \spcend ,
\end{equation}
where
\begin{equation*}
  \label{eqn:filvara}
  \filvara =
  \sumlmb
  \noisecl_\el^{-1} |\shc{\tmpl}{\el}{m}|^2
\end{equation*}
and $\shc{(\fil_\scale^{\rm MF})}{\el}{m} = \innerp{\fil_\scale^{\rm MF}}{\shf{\el}{\m}}$. Here and subsequently we use the shorthand notation \mbox{$\sumlmb = \sum_{\el=0}^{\elmax} \summ$}.  On inspection of the filtering operation in harmonic space given by \eqn{\ref{eqn:filtering_harmonic}}, the matched filter given by \eqn{\ref{eqn:mf}} is justified intuitively since the filter promotes harmonic modes where the source template $\shc{\tmpl}{\el}{m}$ is large and suppresses modes where the noise power $\noisecl_\el$ is large.  

\begin{figure*}
\centering
\subfigure[$\thetacrit=5\degrees$]{
  \includegraphics[viewport=0 0 220 220, clip=, height=45mm]
  {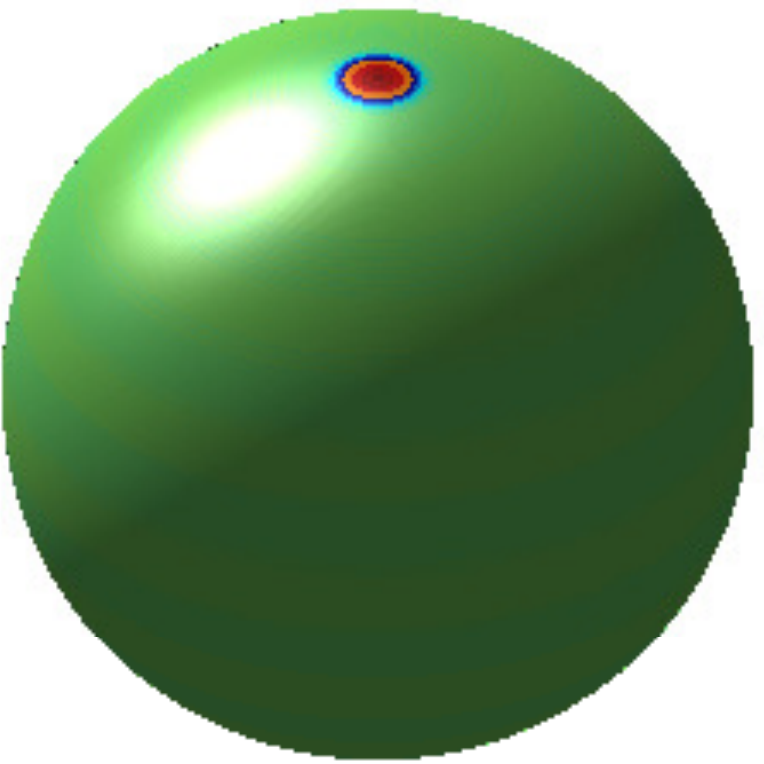}} \quad \quad
\subfigure[$\thetacrit=10\degrees$]{
  \includegraphics[viewport=0 0 220 220, clip=, height=45mm]
  {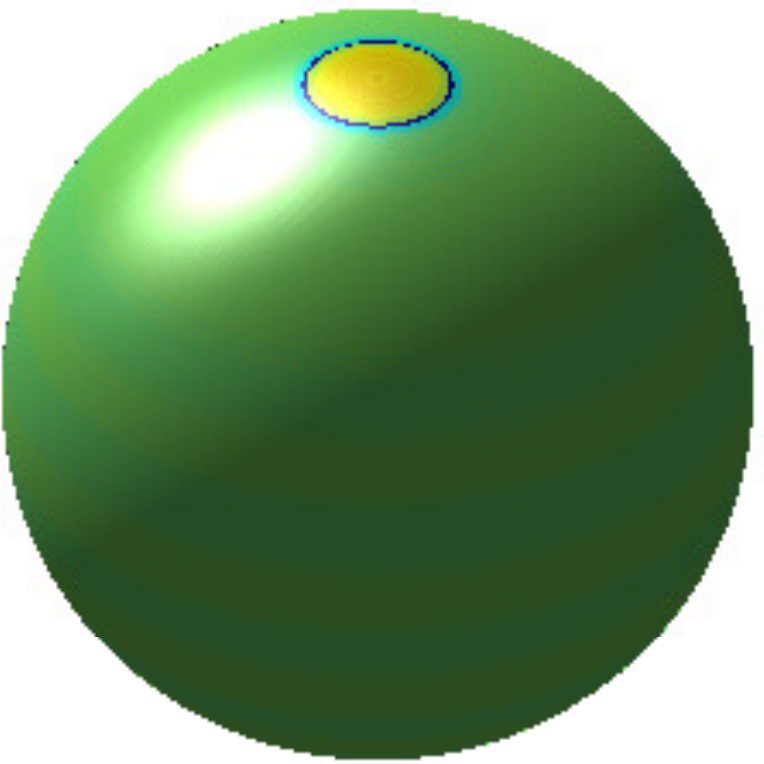}} \quad \quad
\subfigure[$\thetacrit=20\degrees$]{
  \includegraphics[viewport=0 0 220 220, clip=, height=45mm]
  {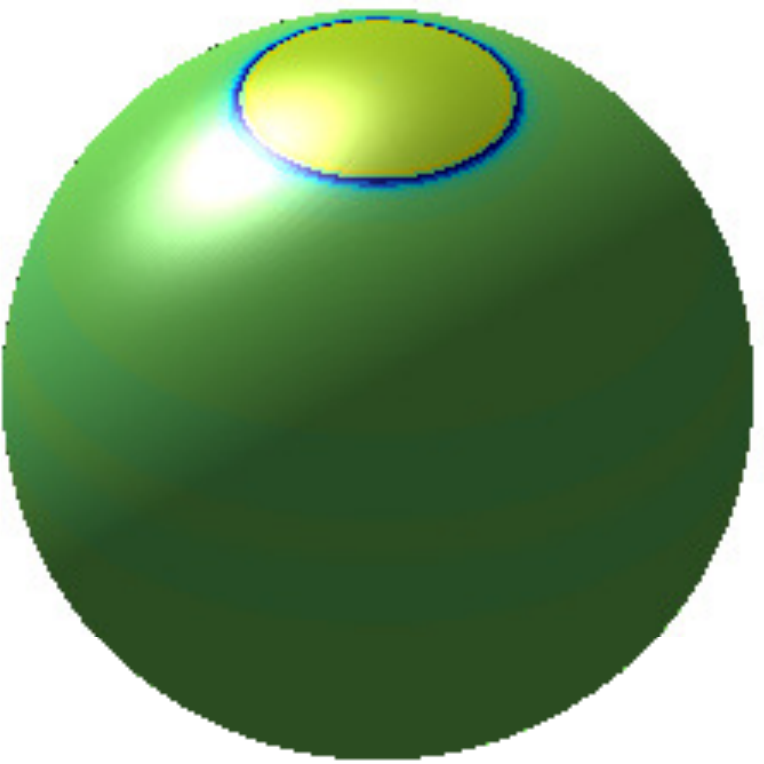}}
\subfigure[$\thetacrit=30\degrees$]{
  \includegraphics[viewport=0 0 220 220, clip=, height=45mm]
  {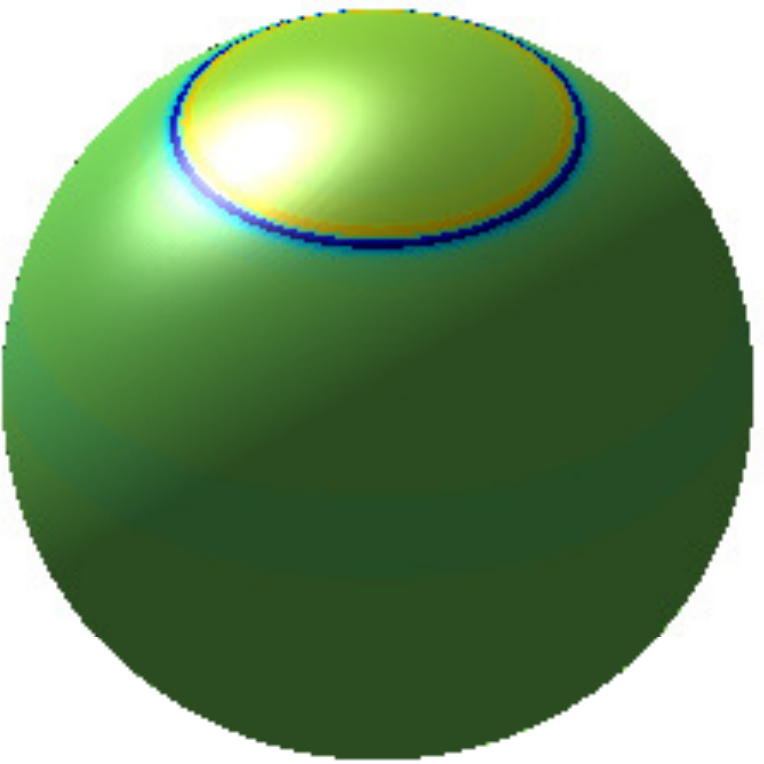}} \quad \quad
\subfigure[$\thetacrit=60\degrees$]{
  \includegraphics[viewport=0 0 220 220, clip=, height=45mm]
  {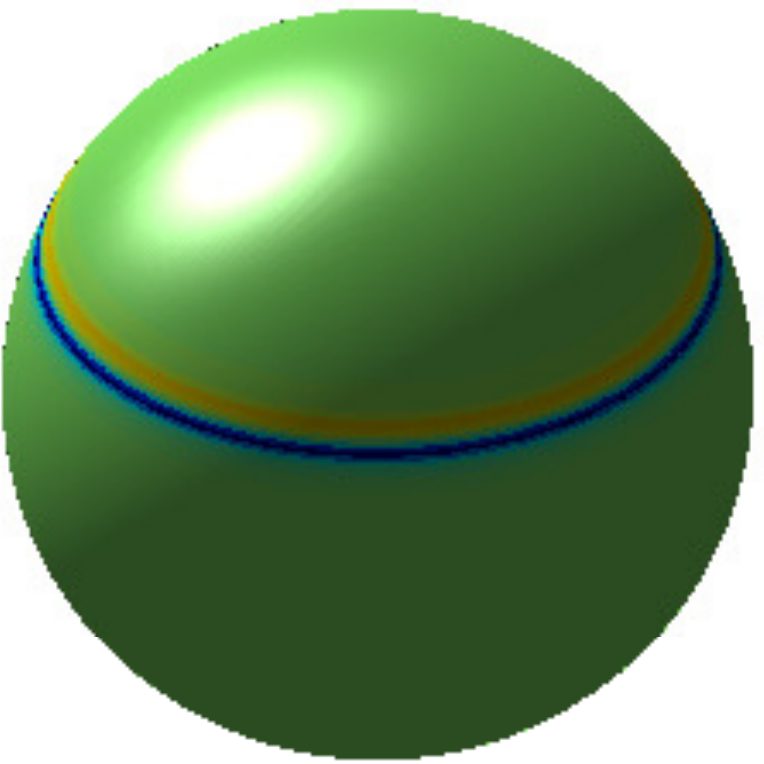}} \quad \quad
\subfigure[$\thetacrit=90\degrees$]{
  \includegraphics[viewport=0 0 220 220, clip=, height=45mm]
  {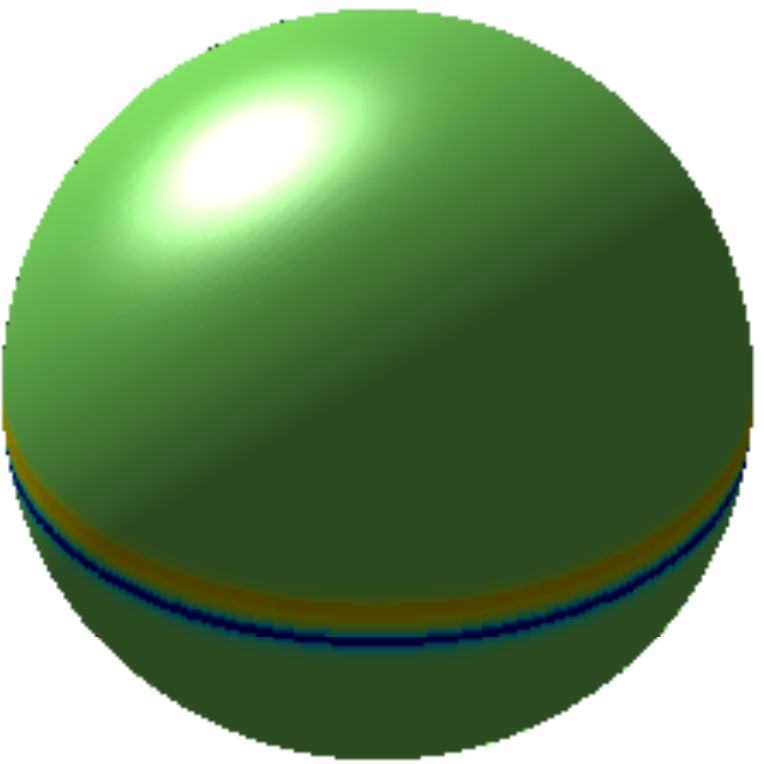}}
\caption{Matched filters optimized to bubble collision signatures of varying size embedded in a $\Lambda$CDM \cmb\ background defined by the best-fit WMAP7$+$BAO$+$H0 power spectrum.}
\label{fig:filters}
\end{figure*}

In \fig{\ref{fig:filters}} we plot the matched filters that are optimized to bubble collision signatures of varying size embedded in a \cmb\ background defined by the $\Lambda$CDM power spectrum that best fits \wmap\ \mbox{7-year}, baryon acoustic oscillations and supernovae observations (hereafter we refer to this spectrum as the best-fit WMAP7$+$BAO$+$H0 power spectrum) \cite{larson:2011}.  Notice that on smaller scales the matched filter contains a central broad hot region to enhance the main bubble collision contribution, surrounded by hot and cold rings to enhance the collision edge. However, on larger scales notice that the matched filter contains only the hot and cold rings that enhance the collision edge.  Since the CMB has more power on large scales, the matched filters on large scales do not look for the large-scale features of the bubble collision signature but rather the transition region near the location where the template goes to zero. Note that the transition region is the best place to look even though the matched filter is constructed for templates with $z_{\rm crit} \sim 0\:\mu{\rm \kelvin}$.

Alternative optimal filters have also been proposed, such as the scale-adaptive filter, defined in Euclidean space by \refns{\cite{sanz:2001,herranz:2002}} and extended to the sphere by \refns{\cite{schaefer:2004,mcewen:2006:filters}}.  Like the matched filter, the scale-adaptive filter minimizes the variance of the filtered field while still providing an unbiased estimate of the source amplitude, but it also imposes a local peak in the filtered field over scale \scale.  Since an additional constraint is imposed when solving the optimization problem that defines the scale-adaptive filter, the \snr\ for the scale-adaptive filter is inevitably lower than that for the matched filter.  However, in the case of (i) a scale-invariant background and (ii) a template profile that changes size simply through a scaling of \saa, the peak in the scale-adaptive filter field can help to find sources of unknown size.  When criteria (i) and (ii) hold, the scale-adaptive filter for a given source size can be constructed by scaling the scale-adaptive filter for a source of a different size.  A filter of incorrect size (since the underlying size of the source is unknown), and scaled variants of it, may then be applied; the peak imposed in scale when constructing the filter can then be used to estimate the unknown source size.  However, neither criterion holds for the case of bubble collision signatures embedded in the \cmb.  Furthermore, although the scale-adaptive filter has been derived on the sphere by
 \refns{\cite{schaefer:2004,mcewen:2006:filters}}, small-angle approximations are made in these derivations; hence the scale-adaptive filter constraints may break down for sources of very large size, such as the bubble collision signatures of interest.  Indeed, we have performed numerical experiments that have shown this to be the case.  Consequently, we do not consider the scale-adaptive filter further.  The problem of detecting sources of unknown size is considered further in \sectn{\ref{sec:filters:algorithm}}.

\subsubsection{Signal-to-noise ratio comparison}

\begin{figure}
\centering
\subfigure[Known source size]{\includegraphics[height=60mm]
  {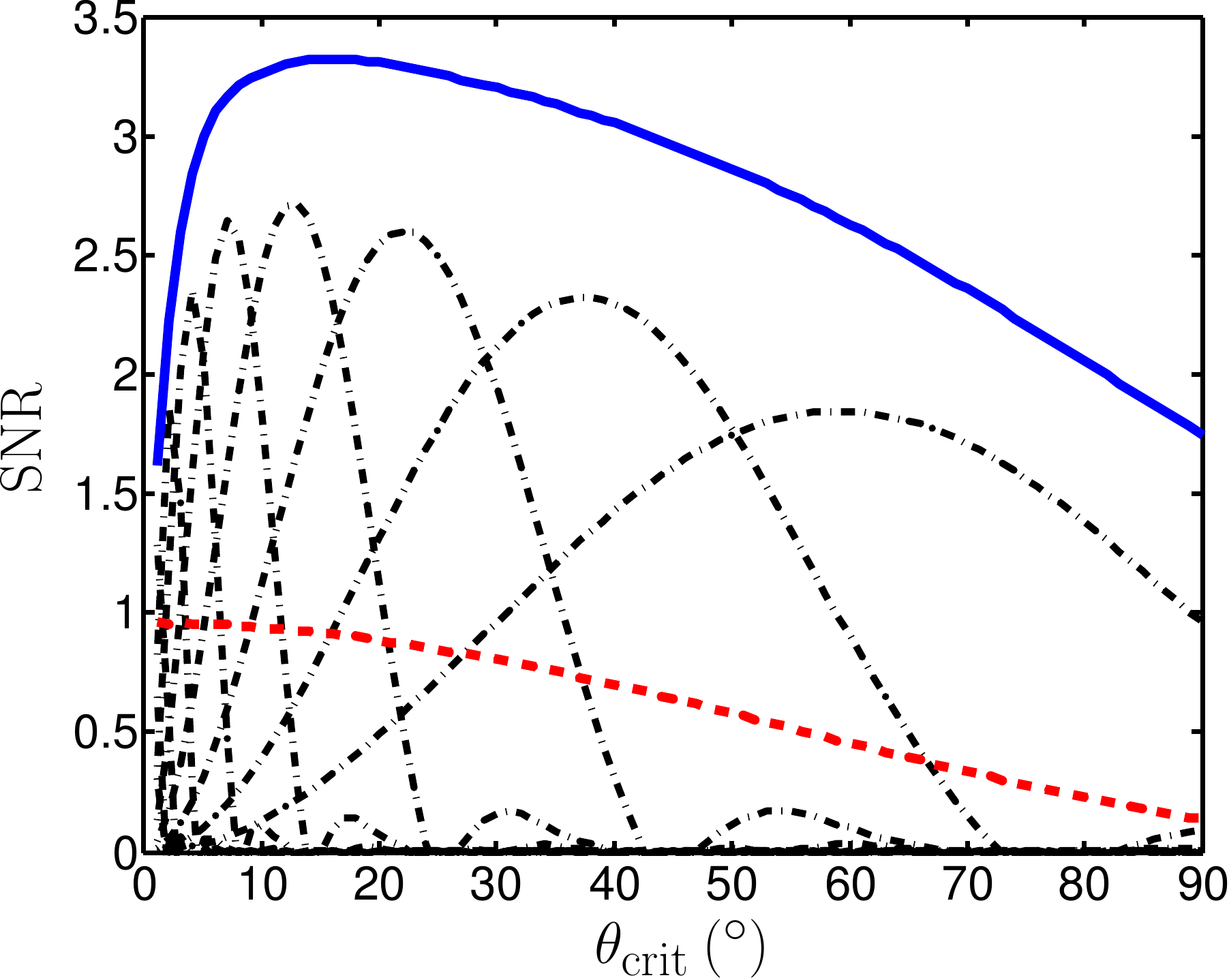}}
\subfigure[Unknown source size]{\includegraphics[height=60mm]
  {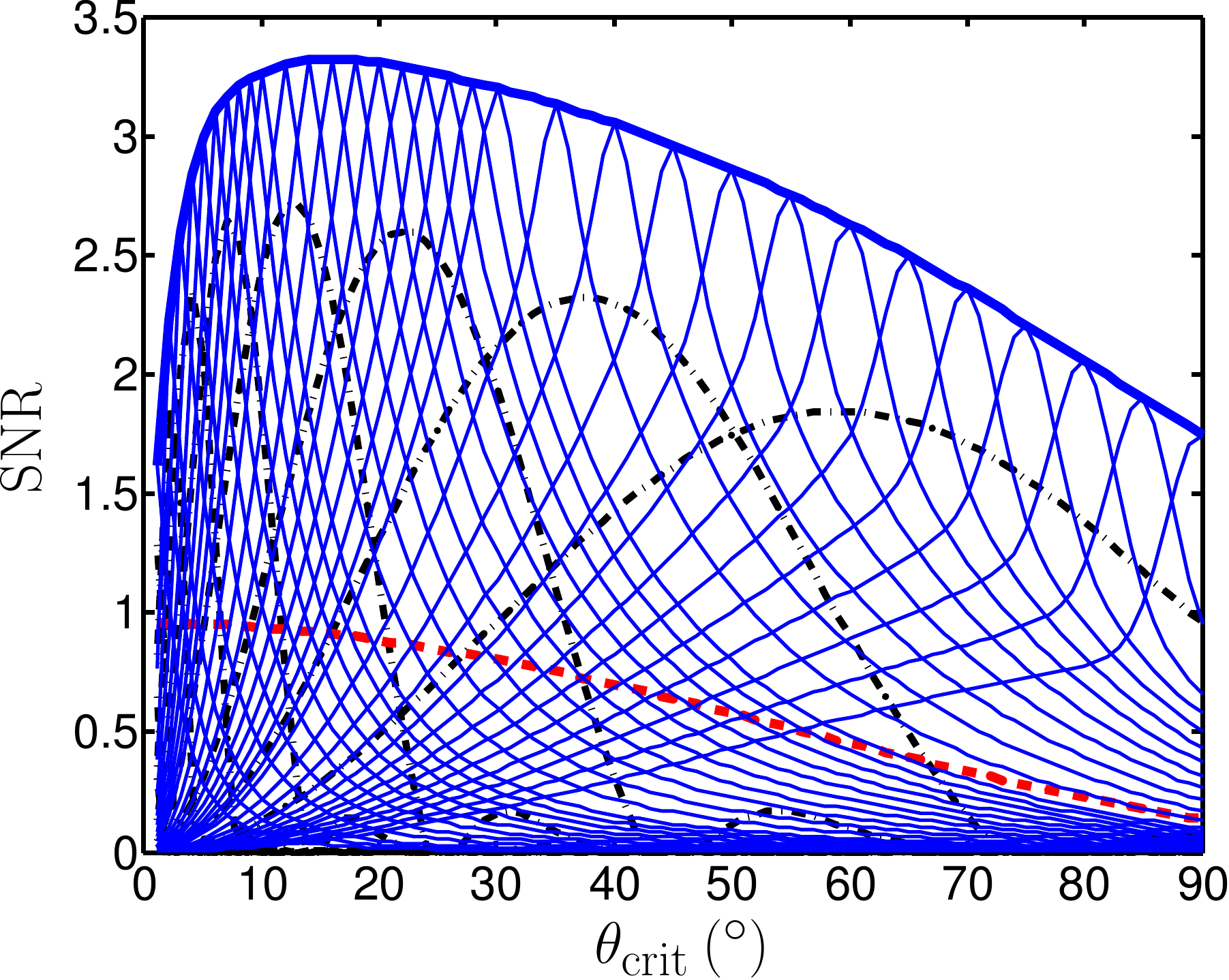}}
\caption{\snr s of bubble collision signatures of varying size with amplitude $\amp=\zo=100\ \mu$\kelvin\ embedded in a $\Lambda$CDM \cmb\ background defined by the best-fit WMAP7$+$BAO$+$H0 power spectrum.  \snr\ curves are plotted for matched filters (solid blue curve), needlets with scaling parameter $B=1.8$ for a range of needlet scales $j$ (dot-dashed black curves) and for the unfiltered field (dashed red curve).  Notice the clear superiority of the matched filter.  In panel~(b) \snr\ curves for the matched filters constructed at a given scale and applied at all other scales are also shown (light solid blue curves).  The scale for which the filters are constructed may be read off the plot from the intersection of the heavy and light solid blue curves.  Provided the \thetacrit\ grid is sampled sufficiently densely, the matched filters remain superior to needlets.}
\label{fig:snr}
\end{figure}

We compare the \snr\ for the matched filter, which by definition is optimal, with the \snr\ for needlets and the unfiltered field.  For a arbitrary filter $\fil_\scale$, such as needlets, the \snr\ defined by \eqn{\ref{eqn:snr}} becomes
\begin{equation*}
\detect^{\fil}_\scale
=
\frac{\amp
\sumlmb
\shc{\tmpl}{\el}{m}
\shc{(\fil_\scale)}{\el}{m}^\cconj
}
{\sqrt{\sumlmb
\noisecl_\el
\bigl | \shc{(\fil_\scale)}{\el}{m} \bigr |^2}
}
\spcend ,
\end{equation*}
where $\shc{(\fil_\scale)}{\el}{m} = \innerp{\fil_\scale}{\shf{\el}{\m}}$.  For the case of the matched filter this expression reduces to \cite{mcewen:2006:filters}
\begin{equation*}
\detect^{\rm \mf}_\scale
=
\filvara^{1/2} \: \amp
\spcend .
\end{equation*}
Finally, we also consider the \snr\ of the unfiltered field, defined by the ratio of its mean and dispersion at the location of a source, given by
\begin{equation*}
\detect^{\rm orig}
=
\frac{\amp
\sumlmb
\sqrt{\frac{2\el+1}{4\pi} \frac{(\el-\m)!}{(\el+\m)!}} \:
\shc{\tmpl}{\el}{m}
}
{\sqrt{\sum_{\el}
\frac{2\el+1}{4\pi} \noisecl_\el
}}
\spcend .
\end{equation*}
The \snr s computed for bubble collision signatures of varying size embedded in a \cmb\ background defined by the $\Lambda$CDM best-fit WMAP7$+$BAO$+$H0 power spectrum are plotted in \fig{\ref{fig:snr}~(a)}.  Notice the superiority of the matched filter to both needlets and the original unfiltered field.

\subsection{Candidate bubble collision detection}
\label{sec:filters:algorithm}

We have selected the optimal matched filter as the filter of choice, since the matched filter optimizes the \snr\ of the filtered field at the position of a source, but thus far we have only considered source profiles of known size.  Here we describe an algorithm using the matched filter to detect multiple sources of unknown and differing size.  The algorithm proceeds as follows.

\begin{enumerate}

\item
Construct matched filters optimized to the source signatures for a grid of scales, \ie\ \mbox{$\scale \in \{ \thetacrit^k \}_{k=1}^{N_{\thetacrit}}$}.

\item
Filter the sky with the matched filter for each scale 
$\scale$.

\item
Compute significance maps 
\begin{equation}
\label{eqn:sig}
S_\scale(\sas) = 
\frac{| \filcoeff_\scale(\sas) - \mu_\scale(\sas) |}
{\sigma_\scale(\sas)}
\spcend ,
\end{equation}
for each filter scale $\scale$.
The mean and dispersion of the filtered field is computed over realisations of the noise process in the absence of sources.

\item
Threshold the significance maps for each filter scale $\scale$,
setting all values of $S_\scale(\sas) < N_{\sigma_\scale}$ to zero. 

\item
Find localized peaks in the thresholded significance maps for each filter scale \scale\ and associate each with a potential detection of a source.

\item 
For each potential detection at a given scale \scale, look across adjacent scales \mbox{$\scale_{\rm adj} \in \{ \scale_{\rm adj } \in \{ \thetacrit^k \}_{k=1}^{N_{\thetacrit}}:$} \mbox{$| \scale_{\rm adj} - \scale | \leq \theta_{\rm adj} \}$} and eliminate the potential detection if a stronger potential detection is made on an adjacent scale.  Potential detections are eliminated as follows.  If adjacent scales contain an overlapping non-zero thresholded region, and if the pixel with the maximum absolute value of the filtered field in the thresholded region is the same sign as the corresponding value at the current scale, but greater in magnitude, then discard the potential detection at the current scale.  Otherwise retain the potential detection and classify it as a detected source.

\item 
For all detected sources, estimate the parameters of the source size, location and amplitude, using the corresponding filter scale, peak position of the thresholded significance map and amplitude of the filtered field, respectively.

\end{enumerate}
The construction of optimal filters is implemented in the \stwofil\ code \cite{mcewen:2006:filters} (which in turn relies on the codes \stwo\ \cite{mcewen:2006:fcswt} and \healpix\ \cite{gorski:2005}), while the \comb\ code  \cite{mcewen:2006:filters} has been used to simulate bubble collisions signatures embedded in a \cmb\ background.\footnote{\stwofil, \stwo\ and \comb\ are available from \url{http://www.jasonmcewen.org/}, while \healpix\ is available from \url{http://healpix.jpl.nasa.gov/}.}  The candidate object detection algorithm described here is implemented in a modified version of \stwofil\ that will soon be made publicly available.

\begin{figure}
\centering
\includegraphics[height=60mm]
  {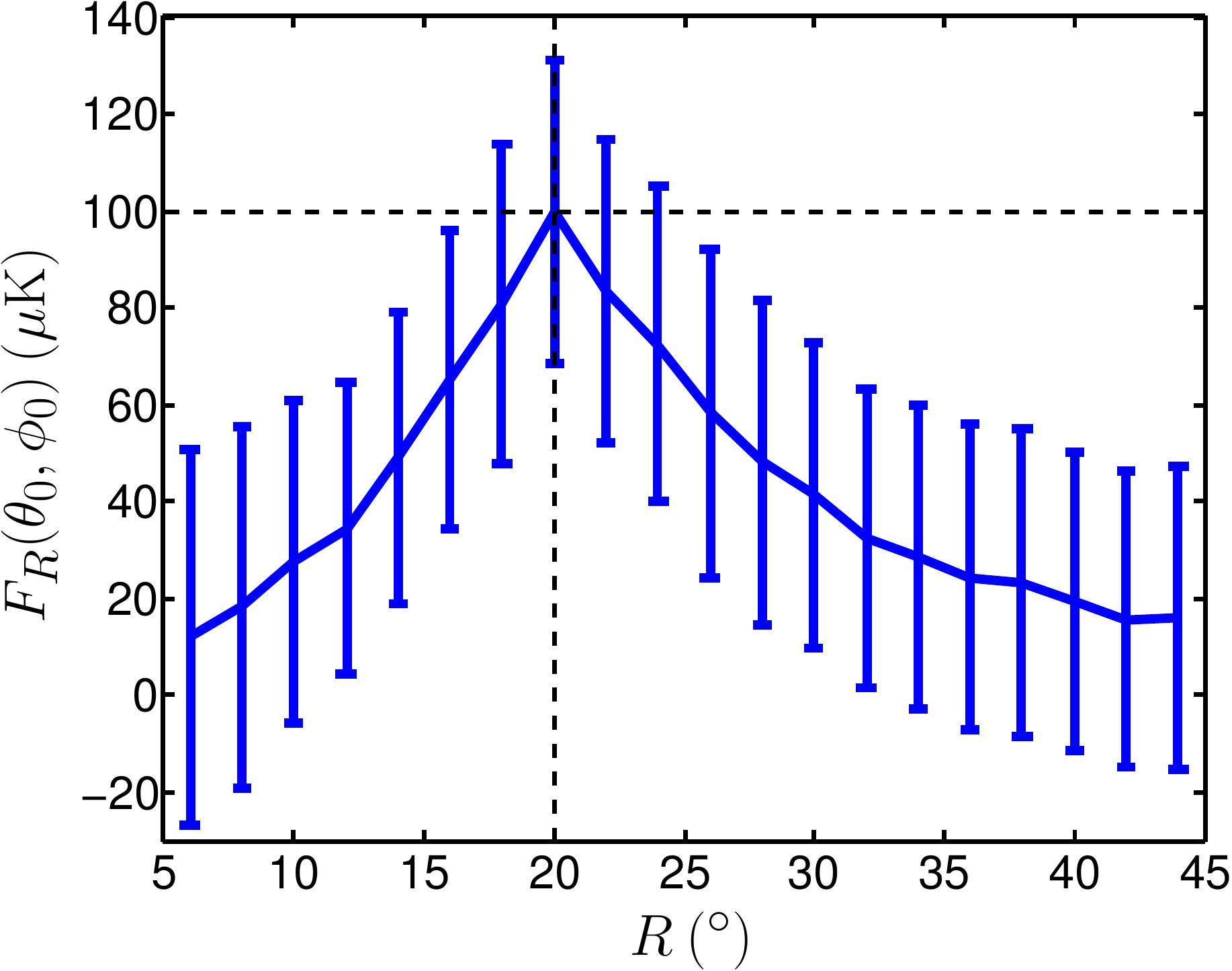}
\caption{Amplitude of the filtered field at the position of a bubble collision signature versus the scale used to construct the corresponding matched filter.  The underlying bubble collision signature has parameters $\{\zo,\thetacrit,\saso \}=\{ 100\ \mu{\rm \kelvin, 20\degrees, 0\degrees, 0\degrees}\}$ and is embedded in a $\Lambda$CDM \cmb\ background defined by the best-fit WMAP7$+$BAO$+$H0 power spectrum.  The solid curve shows the mean value obtained over 100 \cmb\ realizations, while the error bars show the corresponding standard deviation.  Notice that a peak is clearly visible at the scale of the underlying bubble collision signature.  Furthermore, the amplitude of the filtered field at the source scale gives an unbiased estimate of the collision amplitude, as imposed through the construction of the matched filter.}
\label{fig:AvsThetac}
\end{figure}

There is no theoretical guarantee that the peak in the filtered field across scales will coincide with the scale of the unknown source.  Nevertheless, for bubble collision signatures embedded in the \cmb\ we have found, through numerical simulations, that there is indeed such a peak at the scale of an underlying source, as illustrated in \fig{\ref{fig:AvsThetac}}.  Thus, the algorithm outlined above is an effective approach to detecting multiple bubble collision signatures of unknown and differing size.  In situations where a peak does not occur at the scale of an underlying source, numerical simulations may be performed to fit the curve of the filtered field across scales to an underlying source size.  The algorithm outlined above would therefore remain applicable, with only minor alterations.

Although this algorithm considers a grid of candidate scales \mbox{$\scale \in \{ \thetacrit^k \}_{k=1}^{N_{\thetacrit}}$}, it is likely that a source may exist at scales between the samples of the grid.  It is thus important to examine how sensitive the matched filter is to small errors in the source size.  In \fig{\ref{fig:snr}~(b)} we plot \snr\ curves for matched filters constructed on the grid of candidate scales for bubble collision signatures embedded in the \cmb.  A degradation in the \snr\ away from the scale used to construct each filter is clearly apparent; however, provided that the \thetacrit\ grid is sampled sufficiently densely, the matched filters remain effective and are superior to needlets.

The algorithm described above has just two parameters.  The first is the distance $\theta_{\rm adj}$ for which scales are considered to be adjacent, which can be set relative to the grid of candidate sizes.  The second parameter is the threshold level $N_{\sigma_\scale}$, which may be allowed to vary for each filter scale \scale.  The threshold levels may be calibrated from simulations in order to allow a manageable number of false detections, while remaining sensitive to weak source signals.

\section{Bubble collision candidates in WMAP \mbox{7-year} observations}
\label{sec:wmap}

In this section we apply the optimal-filter-based source detection algorithm described in \sectn{\ref{sec:filters}} to \wmap\ \mbox{7-year} observations of the \cmb\ to search for signatures of bubble collisions.  Firstly, we construct optimal filters matched to \wmap\ observations and then calibrate the detection algorithm on a realistic \wmap\ end-to-end simulation.  We then study the sensitivity of the optimal-filter-based detection algorithm.  Finally, we apply the algorithm to \wmap\ observations, resulting in the detection of a number of new candidate bubble collision signatures.

\subsection{Optimal bubble collision filters for \wmap}

We analyze foreground-cleaned \wmap\ \mbox{7-year} W-band observations since this band has the highest resolution beam (with full-width-half-maximum ${\rm \fwhm}=13.2$ arcmin) and suffers from the least foreground contamination \cite{jaroski:2010}.  We restrict our analysis to the band-limit $\elmax=256$ since this is sufficient to represent the bubble collision signatures of interest, which are relatively large scale.  The stochastic background in which the bubble collision signatures live, and that is used to derive matched filters, is defined by the \cmb\ power spectrum, where we assume the best-fit \wmap7$+$BAO$+$H0 best-fit $\Lambda$CDM power spectrum.  The noise considered in the derivation of the matched filter is assumed to be homogenous and isotropic, whereas \wmap\ observations exhibit anisotropic noise that varies over the sky. We therefore neglect \wmap\ noise when constructing optimal filters.  This approximation is valid since the W-band instrumental noise is sub-dominant relative to the \cmb\ contribution in the harmonic region of interest ($\elmax\leq256$).\footnote{We have tested the validity of this assumption by successfully detecting synthetic bubble collision signatures embedded in simulated \wmap\ observations that do include anisotropic noise.}

The optimal filters matched to \wmap\ W-band observations are then computed by \eqn{\ref{eqn:mf}}, where the noise power spectrum $\noisecl_\el$ is given by the \cmb\ spectrum, and the harmonic coefficients of the normalized template profile $\shc{\tmpl}{\el}{\m}$ are modulated by the Legendre coefficients of an azimuthally-symmetric Gaussian beam with ${\rm \fwhm}=13.2$ arcmin. The matched filters computed in this setting are very similar to those displayed in \fig{\ref{fig:filters}}, that were computed in the absence of a beam.\footnote{The Gaussian beam employed in this work is an approximation to the true W-band beam~\cite{jaroski:2010}. As the matched filters computed in the absence of a beam are very similar to those computed with a Gaussian beam, any effects due to the approximated beam are negligible.}

For the algorithm to detect candidate bubble collision signatures of unknown and varying size described in \sectn{\ref{sec:filters:algorithm}}, we must construct matched filters for a grid of scales.  We consider the scales
$\scale \in \{$
$1\degrees$,
$1.5\degrees$,
$2\degrees$,
$3\degrees$,
$4\degrees$,
$5\degrees$,
$6\degrees$,
$7\degrees$,
$8\degrees$,
$9\degrees$,
$10\degrees$,
$12\degrees$,
$14\degrees$,
$16\degrees$,
$18\degrees$,
$20\degrees$,
$22\degrees$,
$24\degrees$,
$26\degrees$,
$28\degrees$,
$30\degrees$,
$35\degrees$,
$40\degrees$,
$45\degrees$,
$50\degrees$,
$55\degrees$,
$60\degrees$,
$65\degrees$,
$70\degrees$,
$75\degrees$,
$80\degrees$,
$85\degrees$,
$90\degrees
\}$.  The \snr\ curves for the matched filters constructed for these scales are shown in \fig{\ref{fig:snr}~(b)} (albeit in the absence of a beam, although the \snr\ curves do not change markedly when these effects are included).  This grid of scales is thus sufficiently sampled to ensure that the matched filters remain effective for scales between the samples of the grid.

\subsection{Calibration}

It is necessary to calibrate the optimal-filter-based bubble collision detection algorithm to realistic \wmap\ observations.  Throughout the calibration we apply the \wmap\ KQ75 mask \cite{gold:2011} since we will adopt this conservative mask when analyzing the \wmap\ data.  Firstly, for each scale \scale, we use 3,000 Gaussian CMB \wmap\ simulations with  W-band beam and anisotropic instrumental noise to compute the mean and dispersion of the filtered field in the absence of sources, as required to compute significance maps of each filtered field through \eqn{\ref{eqn:sig}}.  Based on the sampling of the grid of scales we set the adjacency parameter to $\theta_{\rm adj} = 5\degrees$.  We then calibrate the threshold levels $N_{\sigma_\scale}$ for each scale \scale\ from a realistic \wmap\ simulation that does not contain bubble collision signatures. The thresholds are chosen to allow a manageable number of false detections while remaining sensitive to weak bubble collision signatures.  For this calibration we use a complete end-to-end simulation of the \wmap\ experiment provided by the \wmap\ Science Team~\cite{gold:2011}.
The temperature maps in this simulation are produced from a simulated time-ordered data stream, which is processed using the same algorithm as the actual data. The data for each frequency band is obtained separately from simulated sources including diffuse Galactic foregrounds, CMB fluctuations, realistic noise, smearing from finite integration time, finite beam size, and other instrumental effects.  We use the foreground-reduced W-band simulation for calibration.  
The threshold levels $N_{\sigma_\scale}$ are selected to allow at most two false detections on each scale on this simulated map (recall that detections on one scale can be eliminated by stronger detections made on adjacent scales).  When running the fully-calibrated candidate bubble collision detection algorithm on the \wmap\ W-band end-to-end simulation, 13 false detections are made (note that this is an identical number of false detections to that obtained using needlets \cite{feeney:2011a,feeney:2011b}).  Processing a single map through the algorithm, including filtering at all 33 scales, requires on the order of seconds on a standard desktop computer.

\subsection{Sensitivity}

\begin{figure}
\centering
\includegraphics[height=60mm]
  {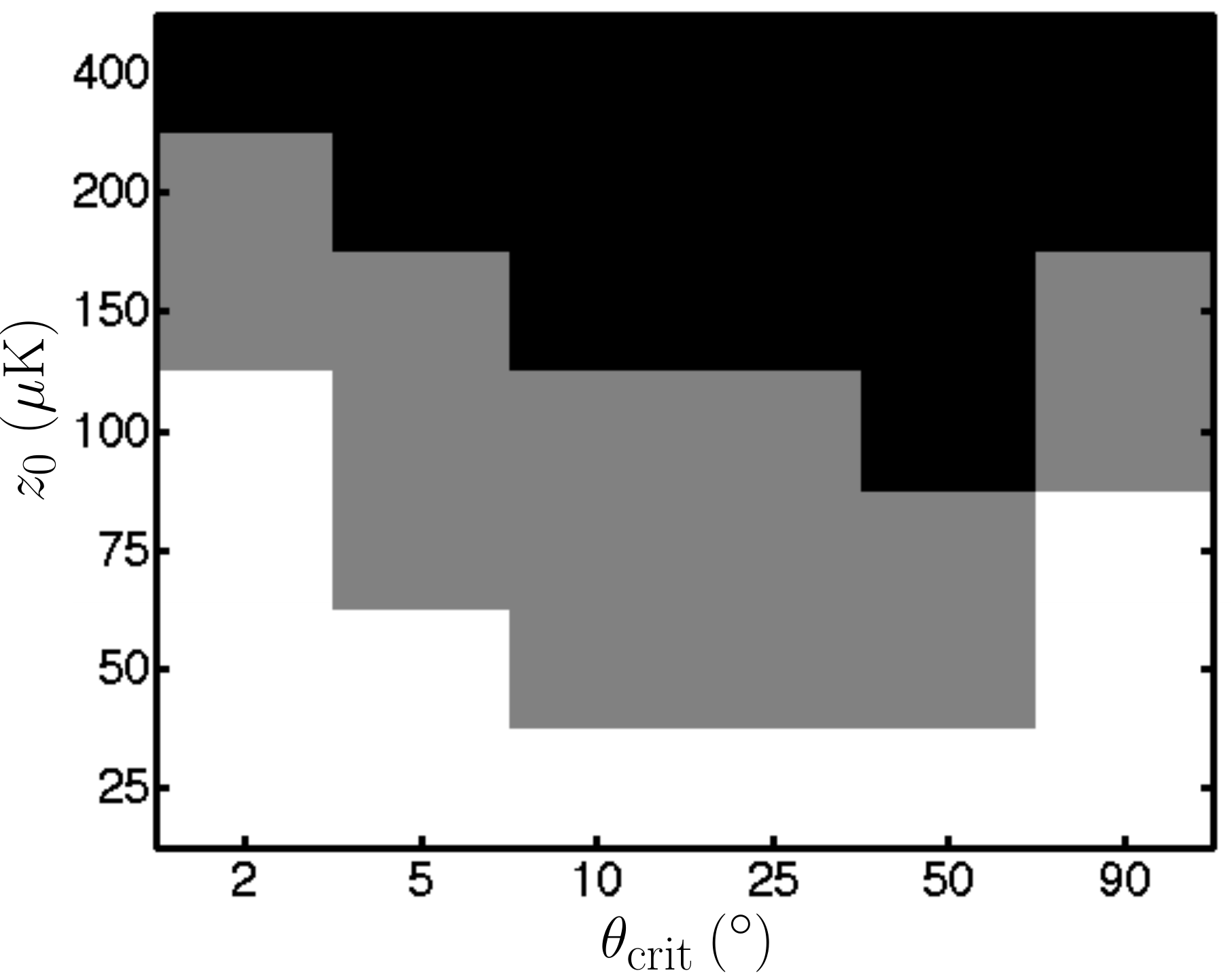}
\caption{Exclusion (black) and sensitivity (grey) regions for the optimal-filter-based bubble collision detection algorithm.  Bubble collision signatures that lie in exclusions regions would certainly be detected by the algorithm provided they were not significantly masked, while collision signatures that lie in sensitivity regions would be detected if they were in a favorable location on the sky.}
\label{fig:exclusion_regions}
\end{figure}

Before applying the calibrated candidate bubble collision detection algorithm to \wmap\ observations, we first assess its sensitivity by applying it to simulated observations where bubble collision signatures are present.  We repeat the sensitivity analysis performed by \refns{\cite{feeney:2011a,feeney:2011b}}, where we lay down known bubble collision signatures on 
low-noise and high-noise regions of the sky, given by locations
$(\saso) = (57.7\degrees,99.2\degrees)$
and
$(\saso) = (56.6\degrees,193.0\degrees)$
respectively, where here and subsequently we adopt Galactic coordinates.  For each collision scale and amplitude that we consider, in each of the low-noise and high-noise regions, we simulate three Gaussian \cmb\ WMAP W-band observations.  We then run the calibrated bubble collision detection algorithm on these six simulations.  If the underlying bubble collision signature is detected in all simulations, we classify the amplitude and scale parameter pair as living in an exclusion region.  If the underlying bubble collision is detected in some but not all simulations, we classify the parameter pair as living in a sensitivity region. If the underlying bubble collision is not detected in any simulation, we classify the parameter pair as living in an unprobed region.  These regions describe the sensitivity of the bubble collision detection algorithm and are plotted in \fig{\ref{fig:exclusion_regions}} for a range of scale and amplitude parameter pairs.  

Bubble collision signatures that lie in exclusion regions would certainly be detected by the optimal-filter-based bubble collision detection algorithm provided they were not significantly masked, while collision signatures that lie in sensitivity regions would be detected if they were in a favorable location on the sky.  When compared to the exclusion and sensitivity regions recovered using needlets \cite{feeney:2011a,feeney:2011b}, the regions recovered using optimal filters are extended to lower temperatures by a factor of $\sim1.7$ in \sky\ for scales $\thetacrit\sim10\degrees$ and most likely further for larger scales (note that the regions plotted in \cite{feeney:2011a,feeney:2011b} are for $\sky/T_0$, where $T_0$ is the average temperature of the \cmb, while here they are plotted for $\sky$).  Optimal filters thus provide an enhancement in sensitivity by a factor of approximately two when compared with needlets, in line with expectations from the \snr\ curves plotted in \fig{\ref{fig:snr}}.  This improvement in sensitivity will be important for uncovering the necessarily weak bubble collision signatures that may be embedded in \cmb\ observations.

\subsection{Candidate bubble collisions}

\begin{figure}
\centering
\subfigure[\wmap\ \mbox{7-year} W-band observations]{
  \parbox{85mm}{\includegraphics[viewport=0 35 800 440,clip=,width=85mm]{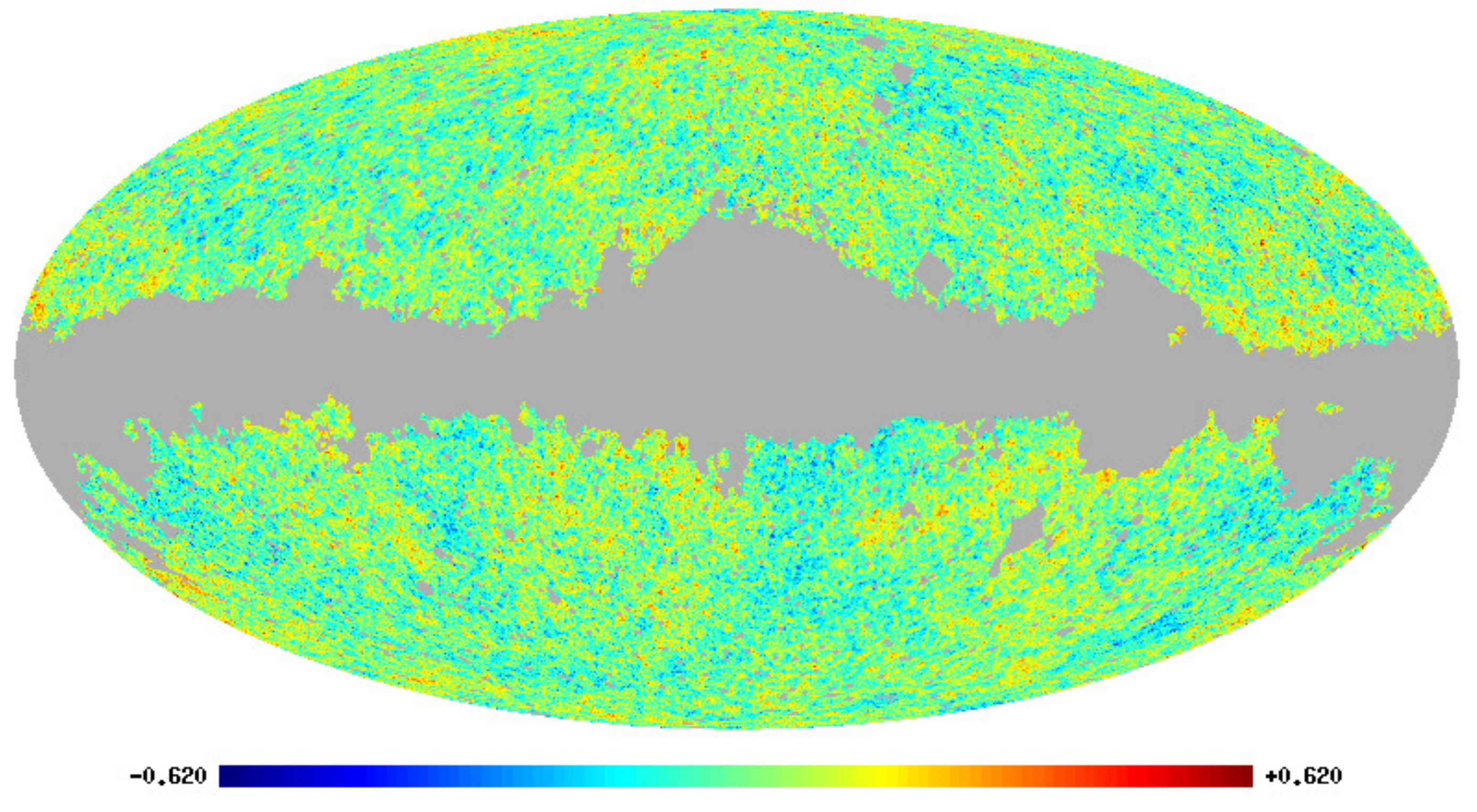}
  \includegraphics[viewport=70 0 730 25,clip=,width=85mm,height=4mm]{figures/wmap_band_imap_r9_7yr_W_v4_maskKQ75_display_scaled}}
}
\subfigure[Candidate bubble collision signatures]{
  \parbox{85mm}{\includegraphics[viewport=0 35 800 440,clip=,width=85mm]{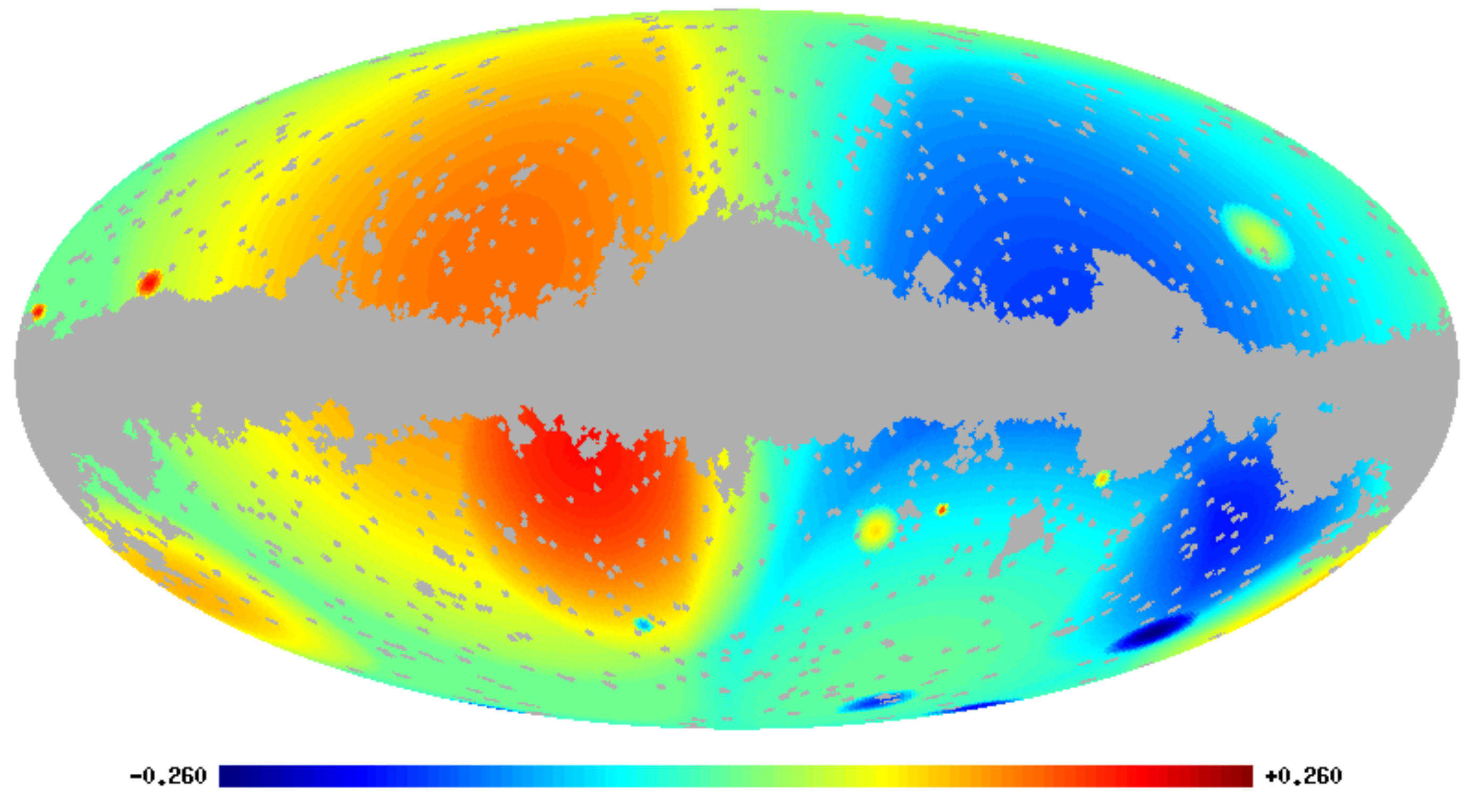}
  \includegraphics[viewport=70 0 730 25,clip=,width=85mm,height=4mm]{figures/out_axiloc_sources_maskKQ75_display_scaled}}
}
\subfigure[Labelled candidate bubble collision signatures]{
  \parbox{85mm}{\includegraphics[viewport=0 35 800 440,clip=,width=85mm]{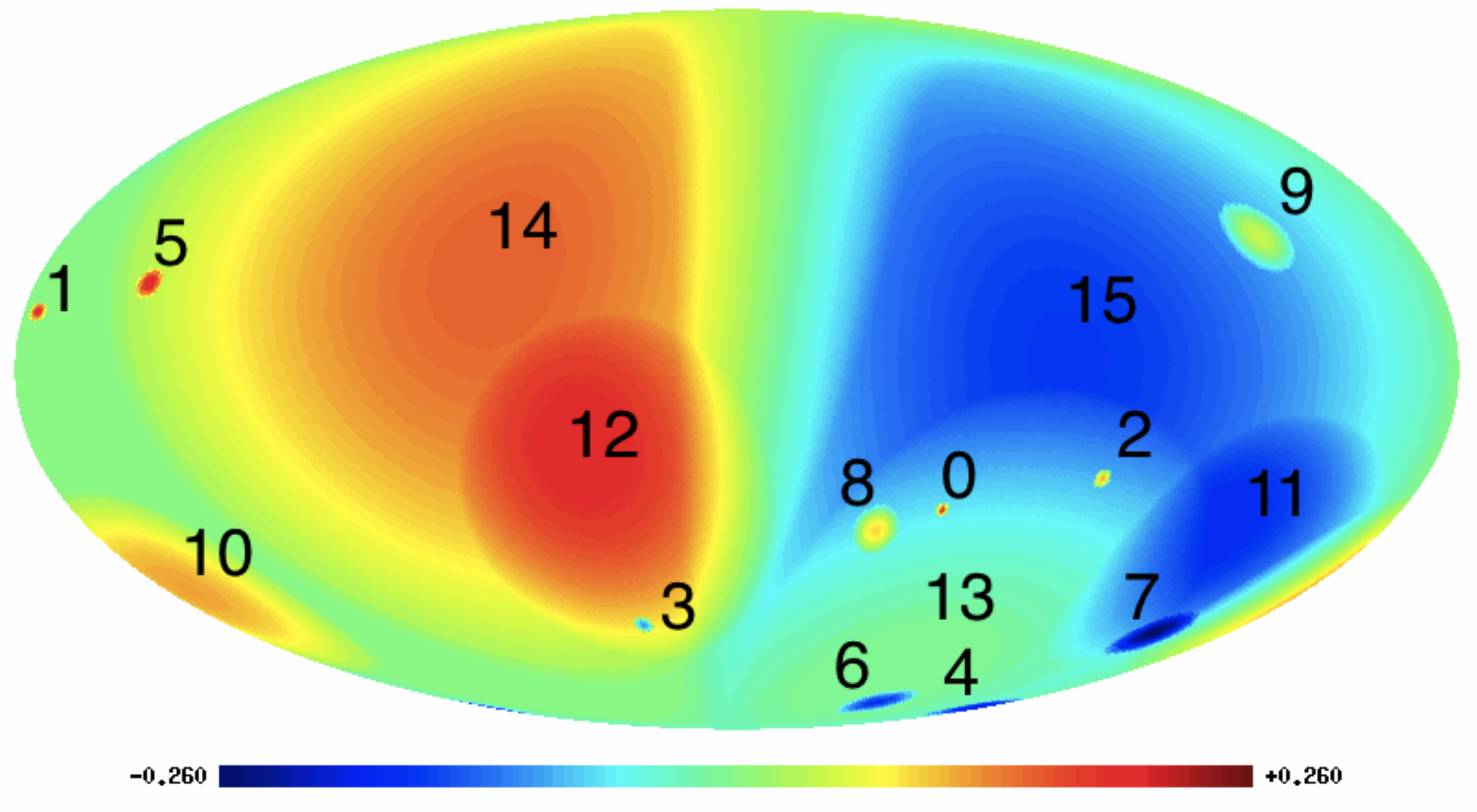}
  \includegraphics[viewport=70 0 730 25,clip=,width=85mm,height=4mm]{figures/out_axiloc_sources_scaled_labelled}}
}
\caption{\wmap\ data analyzed by the bubble collision detection algorithm and the resulting candidate bubble collision signatures detected (in units of m\kelvin).  In panels~(a) and (b) the conservative KQ75 mask is applied. Full-sky maps are plotted using the Mollweide projection.}
\label{fig:wmap}
\end{figure}

The calibrated bubble collision detection algorithm is applied to foreground-cleaned \wmap\ \mbox{7-year} W-band observations \cite{jaroski:2010}, with the conservative KQ75 mask applied \cite{gold:2011}.  16 candidate bubble collision signatures are detected.  The \wmap\ W-band data that are analyzed and the detected bubble collision candidates are plotted on the full-sky in \fig{\ref{fig:wmap}}.  A list of the parameters recovered for each detected candidate is given in \tbl{\ref{tbl:bubble_candidates}}, where the bubble collision candidate labels match those of \fig{\ref{fig:wmap}~(c)}.  In \tbl{\ref{tbl:bubble_candidates}} we also give the significance level of each detection and state whether a feature with similar parameters was detected using needlets \cite{feeney:2011a,feeney:2011b}.  We detect eight new candidate bubble collisions that have not been reported previously.  

As a very preliminary analysis to check that residual foreground contributions are not responsible for the detected candidate bubble collision signatures, we also apply the bubble collision detection algorithm to the foreground-cleaned V-band and Q-band \wmap\ \mbox{7-year} observations. Since foreground contributions are frequency-dependent, one would expect a large difference between the regions detected on different bands if they were due to foreground contributions.  Whether each candidate bubble collision signature is detected in the other \wmap\ bands is listed in the final two columns of \tbl{\ref{tbl:bubble_candidates}}.  All of the new regions detected in the W-band are detected in at least one of the other bands, suggesting residual foregrounds are unlikely to be responsible for the new bubble collision candidates that we detect.

Let us remark that the combination of bubble collision candidates with labels 14 and 15 look somewhat like a dipole contribution.  However, this resemblance is likely to be a coincidence: we know that the matched filters on these large scales enhance ring-like features (see \fig{\ref{fig:filters}}).  Indeed, since the prior on the expected angular size of bubble collision signatures in the CMB is peaked at $90\degrees$~\cite{Freivogel:2009it}, very large candidate bubble collisions are of particular interest. A subsequent Bayesian analysis, following the method of \refns{\cite{feeney:2011a,feeney:2011b}}, will be able to discriminate whether these features are spurious $\Lambda$CDM fluctuations, or else provide evidence for the bubble collision hypothesis.

\begin{table*}
\centering
\caption{Candidate bubble collisions detected in \wmap\ \mbox{7-year} W-band observations.}
\label{tbl:bubble_candidates}
\begin{tabular}{ccccccccc} \hline\hline 
Label \quad\quad & \multicolumn{4}{c}{ Bubble collision parameters} & \quad $S_{\thetacrit}(\saso)$ & \quad  Detected previously \quad & \multicolumn{2}{c}{Detected in other bands} \\\
       & $\zo$ (mK) & $\saao$ (${}^\circ$) & $\sabo$ (${}^\circ$) & $\thetacrit$  (${}^\circ$) &  & & \quad  V-band & Q-band \quad  \\ \hline
    0 &   0.24 & 119.0 & 304.5 &   1.5 & 4.25 & N & Y & N \\
    1 &   0.20 &  78.3 & 176.5 &   2   & 4.15 & N & N & Y \\
    2 &   0.20 & 112.3 & 264.4 &   2   & 4.08 & Y & Y & Y \\
    3 &  -0.19 & 145.1 &  33.0 &   2   & 4.05 & Y & N & N \\
    4 &  -0.17 & 169.0 & 187.5 &   3   & 4.26 & Y & Y & Y \\
    5 &   0.17 &  72.4 & 150.8 &   3   & 4.02 & Y & Y & Y \\
    6 &  -0.16 & 167.2 & 268.7 &   4   & 4.56 & Y & Y & Y \\
    7 &  -0.16 & 147.4 & 207.1 &   5   & 4.67 & Y & Y & Y \\
    8 &   0.15 & 123.2 & 321.3 &   5   & 4.43 & Y & Y & Y \\
    9 &   0.14 &  62.7 & 220.4 &   7   & 4.39 & N & Y & Y \\
   10 &   0.11 & 136.6 & 172.6 &  20   & 3.94 & Y & Y & Y \\
   11 &  -0.09 & 127.2 & 216.9 &  26   & 3.07 & N & N & Y \\
   12 &   0.09 & 116.3 &  31.6 &  35   & 3.33 & N & Y & N \\
   13 &   0.10 & 136.6 & 282.0 &  40   & 3.07 & N & N & Y \\
   14 &   0.15 &  69.6 &  62.6 &  85   & 3.03 & N & Y & N \\
   15 &  -0.16 &  88.5 & 277.7 &  90   & 3.11 & N & Y & Y \\
\hline\hline
\end{tabular}
\end{table*}

\section{Conclusions}
\label{sec:conclusions}

The problem of detecting the existence of a population of sources embedded in the \cmb\ is of widespread interest.  The most unambiguous method of doing so is through a direct evaluation of the full posterior probability distribution of the global parameters of the theory giving rise to the source population.  However, such an approach is computationally impractical for large datasets, such as \wmap\ and Planck.  A method to approximate the full posterior has been developed recently by \refns{\cite{feeney:2011a,feeney:2011b}}.  This approach requires pre-processing of the data to recover a set of candidate sources which are most likely to give the largest contribution to the likelihood.  The pre-processing stage of this method is thus crucial to its overall effectiveness.  Previously needlets were used for candidate source detection \cite{feeney:2011a,feeney:2011b}.  In this article we have developed a new algorithm, based on optimal filtering, to detect candidate sources of unknown and differing angular sizes embedded in full-sky observations of the \cmb.

This method is optimal in the sense that no other filter-based approach can provide a superior enhancement of the source contribution. However, as we have emphasized, the parameters of our algorithm are set to allow some false detections: there is no guarantee that the candidates picked out are the signatures of bubble collisions. The filters will also respond to similar temperature patterns resulting from rare $\Lambda$CDM fluctuations. A further Bayesian model selection step (implementing Occam's razor via a self-consistent penalty for extra model parameters) is required to determine the most likely explanation for the data -- be it a bubble collision, a rare statistical fluctuation of $\Lambda$CDM or something else entirely.

Although our source detection algorithm has general applicability, in this case we have applied it to the problem of detecting candidate bubble collision signatures in \wmap\ \mbox{7-year} observations, where we have demonstrated its superiority.  After calibrating our algorithm on a realistic \wmap\ end-to-end simulation, we have shown both theoretically and through simulations that it provides an enhancement in sensitivity over the previous needlet approach by a factor of approximately two, for an identical number of false detections on the \wmap\ end-to-end simulation.  Applying our algorithm to \wmap\ \mbox{7-year} observations, we detect eight candidate bubble collision signatures that have not been reported previously.  

In a follow-up analysis, we intend to compute the full posterior probability distribution of the number of bubble collision signatures in \wmap\ data using the method developed by \refns{\cite{feeney:2011a,feeney:2011b}}, in light of these new candidate bubble collision signatures.  However, this method was previously restricted to candidate collisions of size $\thetacrit\leq11\degrees$ due to computational memory requirements, while we have detected a number of candidate bubble collision signatures at larger scales.  To handle these large candidate bubble collision signatures, an adaptive-resolution refinement of the method has been developed which processes each candidate at the highest resolution possible given its size and the available computational resources. It was previously shown that the  \wmap\ \mbox{7-year} data do not warrant augmenting $\Lambda$CDM with bubble collisions~{\cite{feeney:2011a,feeney:2011b}. However, the enhanced sensitivity of our optimal-filter-based candidate collision detection algorithm will improve the accuracy of the approximated posterior distribution, and has the potential to uncover evidence for bubble collisions in \wmap\ observations of the \cmb, as well as in next-generation datasets.

\acknowledgements
\ifarxiv
We thank Daniel Mortlock for useful discussions.  SMF thanks David Spergel for an interesting related conversation.  We are very grateful to Eiichiro Komatsu and the \wmap\ Science Team for supplying the end-to-end WMAP simulations used in our null tests. This work was partially supported by a grant from the Foundational Questions Institute (FQXi) Fund, a donor-advised fund of the Silicon Valley Community Foundation on the basis of proposal FQXi-RFP3-1015 to the Foundational Questions Institute. JDM was supported by a Leverhulme Early Career Fellowship from the Leverhulme Trust throughout the completion of this work and is now supported by a Newton International Fellowship from the Royal Society and the British Academy.  SMF is supported by the Perren Fund and STFC. Research at Perimeter Institute is supported by the Government of Canada through Industry Canada and by the Province of Ontario through the Ministry of Research and Innovation. HVP is supported by STFC and the Leverhulme Trust. We acknowledge use of the \healpix\ package and the Legacy Archive for Microwave Background Data Analysis (LAMBDA).  Support for LAMBDA is provided by the NASA Office of Space Science.
\fi

\bibliography{bib}

\begin{thebibliography}{42}%
\makeatletter
\providecommand \@ifxundefined [1]{%
 \@ifx{#1\undefined}
}%
\providecommand \@ifnum [1]{%
 \ifnum #1\expandafter \@firstoftwo
 \else \expandafter \@secondoftwo
 \fi
}%
\providecommand \@ifx [1]{%
 \ifx #1\expandafter \@firstoftwo
 \else \expandafter \@secondoftwo
 \fi
}%
\providecommand \natexlab [1]{#1}%
\providecommand \enquote  [1]{``#1''}%
\providecommand \bibnamefont  [1]{#1}%
\providecommand \bibfnamefont [1]{#1}%
\providecommand \citenamefont [1]{#1}%
\providecommand \href@noop [0]{\@secondoftwo}%
\providecommand \href [0]{\begingroup \@sanitize@url \@href}%
\providecommand \@href[1]{\@@startlink{#1}\@@href}%
\providecommand \@@href[1]{\endgroup#1\@@endlink}%
\providecommand \@sanitize@url [0]{\catcode `\\12\catcode `\$12\catcode
  `\&12\catcode `\#12\catcode `\^12\catcode `\_12\catcode `\%12\relax}%
\providecommand \@@startlink[1]{}%
\providecommand \@@endlink[0]{}%
\providecommand \url  [0]{\begingroup\@sanitize@url \@url }%
\providecommand \@url [1]{\endgroup\@href {#1}{\urlprefix }}%
\providecommand \urlprefix  [0]{URL }%
\providecommand \Eprint [0]{\href }%
\providecommand \doibase [0]{http://dx.doi.org/}%
\providecommand \selectlanguage [0]{\@gobble}%
\providecommand \bibinfo  [0]{\@secondoftwo}%
\providecommand \bibfield  [0]{\@secondoftwo}%
\providecommand \translation [1]{[#1]}%
\providecommand \BibitemOpen [0]{}%
\providecommand \bibitemStop [0]{}%
\providecommand \bibitemNoStop [0]{.\EOS\space}%
\providecommand \EOS [0]{\spacefactor3000\relax}%
\providecommand \BibitemShut  [1]{\csname bibitem#1\endcsname}%
\let\auto@bib@innerbib\@empty
\bibitem [{\citenamefont {{Komatsu}}\ \emph {et~al.}(2011)\citenamefont
  {{Komatsu}}, \citenamefont {{Smith}}, \citenamefont {{Dunkley}},
  \citenamefont {{Bennett}}, \citenamefont {{Gold}}, \citenamefont {{Hinshaw}},
  \citenamefont {{Jarosik}}, \citenamefont {{Larson}}, \citenamefont {{Nolta}},
  \citenamefont {{Page}}, \citenamefont {{Spergel}}, \citenamefont {{Halpern}},
  \citenamefont {{Hill}}, \citenamefont {{Kogut}}, \citenamefont {{Limon}},
  \citenamefont {{Meyer}}, \citenamefont {{Odegard}}, \citenamefont {{Tucker}},
  \citenamefont {{Weiland}}, \citenamefont {{Wollack}},\ and\ \citenamefont
  {{Wright}}}]{komatsu:2010}%
  \BibitemOpen
  \bibfield  {author} {\bibinfo {author} {\bibfnamefont {E.}~\bibnamefont
  {{Komatsu}}}, \bibinfo {author} {\bibfnamefont {K.~M.}\ \bibnamefont
  {{Smith}}}, \bibinfo {author} {\bibfnamefont {J.}~\bibnamefont {{Dunkley}}},
  \bibinfo {author} {\bibfnamefont {C.~L.}\ \bibnamefont {{Bennett}}}, \bibinfo
  {author} {\bibfnamefont {B.}~\bibnamefont {{Gold}}}, \bibinfo {author}
  {\bibfnamefont {G.}~\bibnamefont {{Hinshaw}}}, \bibinfo {author}
  {\bibfnamefont {N.}~\bibnamefont {{Jarosik}}}, \bibinfo {author}
  {\bibfnamefont {D.}~\bibnamefont {{Larson}}}, \bibinfo {author}
  {\bibfnamefont {M.~R.}\ \bibnamefont {{Nolta}}}, \bibinfo {author}
  {\bibfnamefont {L.}~\bibnamefont {{Page}}}, \bibinfo {author} {\bibfnamefont
  {D.~N.}\ \bibnamefont {{Spergel}}}, \bibinfo {author} {\bibfnamefont
  {M.}~\bibnamefont {{Halpern}}}, \bibinfo {author} {\bibfnamefont {R.~S.}\
  \bibnamefont {{Hill}}}, \bibinfo {author} {\bibfnamefont {A.}~\bibnamefont
  {{Kogut}}}, \bibinfo {author} {\bibfnamefont {M.}~\bibnamefont {{Limon}}},
  \bibinfo {author} {\bibfnamefont {S.~S.}\ \bibnamefont {{Meyer}}}, \bibinfo
  {author} {\bibfnamefont {N.}~\bibnamefont {{Odegard}}}, \bibinfo {author}
  {\bibfnamefont {G.~S.}\ \bibnamefont {{Tucker}}}, \bibinfo {author}
  {\bibfnamefont {J.~L.}\ \bibnamefont {{Weiland}}}, \bibinfo {author}
  {\bibfnamefont {E.}~\bibnamefont {{Wollack}}}, \ and\ \bibinfo {author}
  {\bibfnamefont {E.~L.}\ \bibnamefont {{Wright}}},\ }\href@noop {} {\bibfield
  {journal} {\bibinfo  {journal} {Astrophys.J.Suppl.}\ }\textbf {\bibinfo
  {volume} {192}},\ \bibinfo {pages} {18} (\bibinfo {year} {2011})},\ \Eprint
  {http://arxiv.org/abs/arXiv:1001.4538} {arXiv:1001.4538} \BibitemShut
  {NoStop}%
\bibitem [{\citenamefont {{Larson}}\ \emph {et~al.}(2011)\citenamefont
  {{Larson}}, \citenamefont {{Dunkley}}, \citenamefont {{Hinshaw}},
  \citenamefont {{Komatsu}}, \citenamefont {{Nolta}}, \citenamefont
  {{Bennett}}, \citenamefont {{Gold}}, \citenamefont {{Halpern}}, \citenamefont
  {{Hill}}, \citenamefont {{Jarosik}}, \citenamefont {{Kogut}}, \citenamefont
  {{Limon}}, \citenamefont {{Meyer}}, \citenamefont {{Odegard}}, \citenamefont
  {{Page}}, \citenamefont {{Smith}}, \citenamefont {{Spergel}}, \citenamefont
  {{Tucker}}, \citenamefont {{Weiland}}, \citenamefont {{Wollack}},\ and\
  \citenamefont {{Wright}}}]{larson:2011}%
  \BibitemOpen
  \bibfield  {author} {\bibinfo {author} {\bibfnamefont {D.}~\bibnamefont
  {{Larson}}}, \bibinfo {author} {\bibfnamefont {J.}~\bibnamefont {{Dunkley}}},
  \bibinfo {author} {\bibfnamefont {G.}~\bibnamefont {{Hinshaw}}}, \bibinfo
  {author} {\bibfnamefont {E.}~\bibnamefont {{Komatsu}}}, \bibinfo {author}
  {\bibfnamefont {M.~R.}\ \bibnamefont {{Nolta}}}, \bibinfo {author}
  {\bibfnamefont {C.~L.}\ \bibnamefont {{Bennett}}}, \bibinfo {author}
  {\bibfnamefont {B.}~\bibnamefont {{Gold}}}, \bibinfo {author} {\bibfnamefont
  {M.}~\bibnamefont {{Halpern}}}, \bibinfo {author} {\bibfnamefont {R.~S.}\
  \bibnamefont {{Hill}}}, \bibinfo {author} {\bibfnamefont {N.}~\bibnamefont
  {{Jarosik}}}, \bibinfo {author} {\bibfnamefont {A.}~\bibnamefont {{Kogut}}},
  \bibinfo {author} {\bibfnamefont {M.}~\bibnamefont {{Limon}}}, \bibinfo
  {author} {\bibfnamefont {S.~S.}\ \bibnamefont {{Meyer}}}, \bibinfo {author}
  {\bibfnamefont {N.}~\bibnamefont {{Odegard}}}, \bibinfo {author}
  {\bibfnamefont {L.}~\bibnamefont {{Page}}}, \bibinfo {author} {\bibfnamefont
  {K.~M.}\ \bibnamefont {{Smith}}}, \bibinfo {author} {\bibfnamefont {D.~N.}\
  \bibnamefont {{Spergel}}}, \bibinfo {author} {\bibfnamefont {G.~S.}\
  \bibnamefont {{Tucker}}}, \bibinfo {author} {\bibfnamefont {J.~L.}\
  \bibnamefont {{Weiland}}}, \bibinfo {author} {\bibfnamefont {E.}~\bibnamefont
  {{Wollack}}}, \ and\ \bibinfo {author} {\bibfnamefont {E.~L.}\ \bibnamefont
  {{Wright}}},\ }\href {\doibase 10.1088/0067-0049/192/2/16} {\bibfield
  {journal} {\bibinfo  {journal} {Astrophys.\ J.\ Supp.}\ }\textbf {\bibinfo
  {volume} {192}},\ \bibinfo {eid} {16} (\bibinfo {year} {2011})},\ \Eprint
  {http://arxiv.org/abs/1001.4635} {arXiv:1001.4635 [astro-ph.CO]} \BibitemShut
  {NoStop}%
\bibitem [{\citenamefont {Garriga}\ \emph {et~al.}(2007)\citenamefont
  {Garriga}, \citenamefont {Guth},\ and\ \citenamefont
  {Vilenkin}}]{Garriga:2006hw}%
  \BibitemOpen
  \bibfield  {author} {\bibinfo {author} {\bibfnamefont {J.}~\bibnamefont
  {Garriga}}, \bibinfo {author} {\bibfnamefont {A.~H.}\ \bibnamefont {Guth}}, \
  and\ \bibinfo {author} {\bibfnamefont {A.}~\bibnamefont {Vilenkin}},\ }\href
  {\doibase 10.1103/PhysRevD.76.123512} {\bibfield  {journal} {\bibinfo
  {journal} {Phys.\ Rev.\ D.}\ }\textbf {\bibinfo {volume} {D76}},\ \bibinfo
  {pages} {123512} (\bibinfo {year} {2007})},\ \Eprint
  {http://arxiv.org/abs/hep-th/0612242} {arXiv:hep-th/0612242} \BibitemShut
  {NoStop}%
\bibitem [{\citenamefont {Aguirre}\ \emph {et~al.}(2007)\citenamefont
  {Aguirre}, \citenamefont {Johnson},\ and\ \citenamefont
  {Shomer}}]{Aguirre:2007an}%
  \BibitemOpen
  \bibfield  {author} {\bibinfo {author} {\bibfnamefont {A.}~\bibnamefont
  {Aguirre}}, \bibinfo {author} {\bibfnamefont {M.~C.}\ \bibnamefont
  {Johnson}}, \ and\ \bibinfo {author} {\bibfnamefont {A.}~\bibnamefont
  {Shomer}},\ }\href@noop {} {\bibfield  {journal} {\bibinfo  {journal} {Phys.\
  Rev.\ D.}\ }\textbf {\bibinfo {volume} {D76}},\ \bibinfo {pages} {063509}
  (\bibinfo {year} {2007})},\ \Eprint {http://arxiv.org/abs/arXiv:0704.3473
  [hep-th]} {arXiv:0704.3473 [hep-th]} \BibitemShut {NoStop}%
\bibitem [{\citenamefont {Aguirre}\ and\ \citenamefont
  {Johnson}(2008)}]{Aguirre:2007wm}%
  \BibitemOpen
  \bibfield  {author} {\bibinfo {author} {\bibfnamefont {A.}~\bibnamefont
  {Aguirre}}\ and\ \bibinfo {author} {\bibfnamefont {M.~C.}\ \bibnamefont
  {Johnson}},\ }\href {\doibase 10.1103/PhysRevD.77.123536} {\bibfield
  {journal} {\bibinfo  {journal} {Phys.\ Rev.\ D.}\ }\textbf {\bibinfo {volume}
  {D77}},\ \bibinfo {pages} {123536} (\bibinfo {year} {2008})},\ \Eprint
  {http://arxiv.org/abs/0712.3038} {arXiv:0712.3038 [hep-th]} \BibitemShut
  {NoStop}%
\bibitem [{\citenamefont {Aguirre}\ \emph {et~al.}(2009)\citenamefont
  {Aguirre}, \citenamefont {Johnson},\ and\ \citenamefont
  {Tysanner}}]{Aguirre:2008wy}%
  \BibitemOpen
  \bibfield  {author} {\bibinfo {author} {\bibfnamefont {A.}~\bibnamefont
  {Aguirre}}, \bibinfo {author} {\bibfnamefont {M.~C.}\ \bibnamefont
  {Johnson}}, \ and\ \bibinfo {author} {\bibfnamefont {M.}~\bibnamefont
  {Tysanner}},\ }\href {\doibase 10.1103/PhysRevD.79.123514} {\bibfield
  {journal} {\bibinfo  {journal} {Phys.\ Rev.\ D.}\ }\textbf {\bibinfo {volume}
  {D79}},\ \bibinfo {pages} {123514} (\bibinfo {year} {2009})}\BibitemShut
  {NoStop}%
\bibitem [{\citenamefont {Chang}\ \emph {et~al.}(2008)\citenamefont {Chang},
  \citenamefont {Kleban},\ and\ \citenamefont {Levi}}]{Chang:2007eq}%
  \BibitemOpen
  \bibfield  {author} {\bibinfo {author} {\bibfnamefont {S.}~\bibnamefont
  {Chang}}, \bibinfo {author} {\bibfnamefont {M.}~\bibnamefont {Kleban}}, \
  and\ \bibinfo {author} {\bibfnamefont {T.~S.}\ \bibnamefont {Levi}},\ }\href
  {\doibase 10.1088/1475-7516/2008/04/034} {\bibfield  {journal} {\bibinfo
  {journal} {JCAP}\ }\textbf {\bibinfo {volume} {0804}},\ \bibinfo {pages}
  {034} (\bibinfo {year} {2008})}\BibitemShut {NoStop}%
\bibitem [{\citenamefont {Chang}\ \emph {et~al.}(2009)\citenamefont {Chang},
  \citenamefont {Kleban},\ and\ \citenamefont {Levi}}]{Chang:2008gj}%
  \BibitemOpen
  \bibfield  {author} {\bibinfo {author} {\bibfnamefont {S.}~\bibnamefont
  {Chang}}, \bibinfo {author} {\bibfnamefont {M.}~\bibnamefont {Kleban}}, \
  and\ \bibinfo {author} {\bibfnamefont {T.~S.}\ \bibnamefont {Levi}},\ }\href
  {\doibase 10.1088/1475-7516/2009/04/025} {\bibfield  {journal} {\bibinfo
  {journal} {JCAP}\ }\textbf {\bibinfo {volume} {0904}},\ \bibinfo {pages}
  {025} (\bibinfo {year} {2009})},\ \Eprint {http://arxiv.org/abs/0810.5128}
  {arXiv:0810.5128 [hep-th]} \BibitemShut {NoStop}%
\bibitem [{\citenamefont {Czech}\ \emph {et~al.}(2010)\citenamefont {Czech},
  \citenamefont {Kleban}, \citenamefont {Larjo}, \citenamefont {Levi},\ and\
  \citenamefont {Sigurdson}}]{Czech:2010rg}%
  \BibitemOpen
  \bibfield  {author} {\bibinfo {author} {\bibfnamefont {B.}~\bibnamefont
  {Czech}}, \bibinfo {author} {\bibfnamefont {M.}~\bibnamefont {Kleban}},
  \bibinfo {author} {\bibfnamefont {K.}~\bibnamefont {Larjo}}, \bibinfo
  {author} {\bibfnamefont {T.~S.}\ \bibnamefont {Levi}}, \ and\ \bibinfo
  {author} {\bibfnamefont {K.}~\bibnamefont {Sigurdson}},\ }\href {\doibase
  10.1088/1475-7516/2010/12/023} {\bibfield  {journal} {\bibinfo  {journal}
  {JCAP}\ }\textbf {\bibinfo {volume} {1012}},\ \bibinfo {pages} {023}
  (\bibinfo {year} {2010})},\ \Eprint {http://arxiv.org/abs/1006.0832}
  {arXiv:1006.0832 [astro-ph.CO]} \BibitemShut {NoStop}%
\bibitem [{\citenamefont {Dahlen}(2010)}]{Dahlen:2008rd}%
  \BibitemOpen
  \bibfield  {author} {\bibinfo {author} {\bibfnamefont {A.}~\bibnamefont
  {Dahlen}},\ }\href {\doibase 10.1103/PhysRevD.81.063501} {\bibfield
  {journal} {\bibinfo  {journal} {Phys.\ Rev.\ D.}\ }\textbf {\bibinfo {volume}
  {D81}},\ \bibinfo {pages} {063501} (\bibinfo {year} {2010})},\ \Eprint
  {http://arxiv.org/abs/0812.0414} {arXiv:0812.0414 [hep-th]} \BibitemShut
  {NoStop}%
\bibitem [{\citenamefont {Freivogel}\ \emph {et~al.}(2009)\citenamefont
  {Freivogel}, \citenamefont {Kleban}, \citenamefont {Nicolis},\ and\
  \citenamefont {Sigurdson}}]{Freivogel:2009it}%
  \BibitemOpen
  \bibfield  {author} {\bibinfo {author} {\bibfnamefont {B.}~\bibnamefont
  {Freivogel}}, \bibinfo {author} {\bibfnamefont {M.}~\bibnamefont {Kleban}},
  \bibinfo {author} {\bibfnamefont {A.}~\bibnamefont {Nicolis}}, \ and\
  \bibinfo {author} {\bibfnamefont {K.}~\bibnamefont {Sigurdson}},\ }\href
  {\doibase 10.1088/1475-7516/2009/08/036} {\bibfield  {journal} {\bibinfo
  {journal} {JCAP}\ }\textbf {\bibinfo {volume} {0908}},\ \bibinfo {pages}
  {036} (\bibinfo {year} {2009})}\BibitemShut {NoStop}%
\bibitem [{\citenamefont {Larjo}\ and\ \citenamefont
  {Levi}(2010)}]{Larjo:2009mt}%
  \BibitemOpen
  \bibfield  {author} {\bibinfo {author} {\bibfnamefont {K.}~\bibnamefont
  {Larjo}}\ and\ \bibinfo {author} {\bibfnamefont {T.~S.}\ \bibnamefont
  {Levi}},\ }\href {\doibase 10.1088/1475-7516/2010/08/034} {\bibfield
  {journal} {\bibinfo  {journal} {JCAP}\ }\textbf {\bibinfo {volume} {1008}},\
  \bibinfo {pages} {034} (\bibinfo {year} {2010})},\ \Eprint
  {http://arxiv.org/abs/0910.4159} {arXiv:0910.4159 [hep-th]} \BibitemShut
  {NoStop}%
\bibitem [{\citenamefont {Kleban}\ \emph {et~al.}(2011)\citenamefont {Kleban},
  \citenamefont {Levi},\ and\ \citenamefont {Sigurdson}}]{Kleban:2011yc}%
  \BibitemOpen
  \bibfield  {author} {\bibinfo {author} {\bibfnamefont {M.}~\bibnamefont
  {Kleban}}, \bibinfo {author} {\bibfnamefont {T.~S.}\ \bibnamefont {Levi}}, \
  and\ \bibinfo {author} {\bibfnamefont {K.}~\bibnamefont {Sigurdson}},\
  }\href@noop {} {\  (\bibinfo {year} {2011})},\ \Eprint
  {http://arxiv.org/abs/1109.3473} {arXiv:1109.3473 [astro-ph.CO]} \BibitemShut
  {NoStop}%
\bibitem [{\citenamefont {Gobbetti}\ and\ \citenamefont
  {Kleban}(2012)}]{Gobbetti:2012yq}%
  \BibitemOpen
  \bibfield  {author} {\bibinfo {author} {\bibfnamefont {R.}~\bibnamefont
  {Gobbetti}}\ and\ \bibinfo {author} {\bibfnamefont {M.}~\bibnamefont
  {Kleban}},\ }\href@noop {} {\  (\bibinfo {year} {2012})},\ \Eprint
  {http://arxiv.org/abs/1201.6380} {arXiv:1201.6380 [hep-th]} \BibitemShut
  {NoStop}%
\bibitem [{\citenamefont {Aguirre}\ and\ \citenamefont
  {Johnson}(2011)}]{Aguirre:2009ug}%
  \BibitemOpen
  \bibfield  {author} {\bibinfo {author} {\bibfnamefont {A.}~\bibnamefont
  {Aguirre}}\ and\ \bibinfo {author} {\bibfnamefont {M.~C.}\ \bibnamefont
  {Johnson}},\ }\href {\doibase 10.1088/0034-4885/74/7/074901} {\bibfield
  {journal} {\bibinfo  {journal} {Rept.Prog.Phys.}\ }\textbf {\bibinfo {volume}
  {74}},\ \bibinfo {pages} {074901} (\bibinfo {year} {2011})},\ \Eprint
  {http://arxiv.org/abs/0908.4105} {arXiv:0908.4105 [hep-th]} \BibitemShut
  {NoStop}%
\bibitem [{\citenamefont {Vilenkin}\ and\ \citenamefont
  {Shellard}(1986)}]{Vilenkin:1986hg}%
  \BibitemOpen
  \bibfield  {author} {\bibinfo {author} {\bibfnamefont {A.}~\bibnamefont
  {Vilenkin}}\ and\ \bibinfo {author} {\bibfnamefont {E.~P.~S.}\ \bibnamefont
  {Shellard}},\ }\href@noop {} {\emph {\bibinfo {title} {{Cosmic strings and
  other topological defects}}}}\ (\bibinfo  {publisher} {Cambridge University
  Press},\ \bibinfo {year} {1986})\BibitemShut {NoStop}%
\bibitem [{\citenamefont {{Turok}}\ and\ \citenamefont
  {{Spergel}}(1990)}]{turok:1990}%
  \BibitemOpen
  \bibfield  {author} {\bibinfo {author} {\bibfnamefont {N.}~\bibnamefont
  {{Turok}}}\ and\ \bibinfo {author} {\bibfnamefont {D.}~\bibnamefont
  {{Spergel}}},\ }\href@noop {} {\bibfield  {journal} {\bibinfo  {journal}
  {Phys. Rev. Lett.}\ }\textbf {\bibinfo {volume} {64}},\ \bibinfo {pages}
  {2736} (\bibinfo {year} {1990})}\BibitemShut {NoStop}%
\bibitem [{\citenamefont {{Sunyaev}}\ and\ \citenamefont
  {{Zeldovich}}(1980)}]{sz:1980}%
  \BibitemOpen
  \bibfield  {author} {\bibinfo {author} {\bibfnamefont {R.~A.}\ \bibnamefont
  {{Sunyaev}}}\ and\ \bibinfo {author} {\bibfnamefont {I.~B.}\ \bibnamefont
  {{Zeldovich}}},\ }\href {\doibase 10.1146/annurev.aa.18.090180.002541}
  {\bibfield  {journal} {\bibinfo  {journal} {Ann.\ Rev.\ Astron.\ Astrophys.}\
  }\textbf {\bibinfo {volume} {18}},\ \bibinfo {pages} {537} (\bibinfo {year}
  {1980})}\BibitemShut {NoStop}%
\bibitem [{\citenamefont {{Bennett}}\ \emph {et~al.}(2003)\citenamefont
  {{Bennett}}, \citenamefont {{Bay}}, \citenamefont {{Halpern}}, \citenamefont
  {{Hinshaw}}, \citenamefont {{Jackson}}, \citenamefont {{Jarosik}},
  \citenamefont {{Kogut}}, \citenamefont {{Limon}}, \citenamefont {{Meyer}},
  \citenamefont {{Page}}, \citenamefont {{Spergel}}, \citenamefont {{Tucker}},
  \citenamefont {{Wilkinson}}, \citenamefont {{Wollack}},\ and\ \citenamefont
  {{Wright}}}]{bennett:2003c}%
  \BibitemOpen
  \bibfield  {author} {\bibinfo {author} {\bibfnamefont {C.~L.}\ \bibnamefont
  {{Bennett}}}, \bibinfo {author} {\bibfnamefont {M.}~\bibnamefont {{Bay}}},
  \bibinfo {author} {\bibfnamefont {M.}~\bibnamefont {{Halpern}}}, \bibinfo
  {author} {\bibfnamefont {G.}~\bibnamefont {{Hinshaw}}}, \bibinfo {author}
  {\bibfnamefont {C.}~\bibnamefont {{Jackson}}}, \bibinfo {author}
  {\bibfnamefont {N.}~\bibnamefont {{Jarosik}}}, \bibinfo {author}
  {\bibfnamefont {A.}~\bibnamefont {{Kogut}}}, \bibinfo {author} {\bibfnamefont
  {M.}~\bibnamefont {{Limon}}}, \bibinfo {author} {\bibfnamefont {S.~S.}\
  \bibnamefont {{Meyer}}}, \bibinfo {author} {\bibfnamefont {L.}~\bibnamefont
  {{Page}}}, \bibinfo {author} {\bibfnamefont {D.~N.}\ \bibnamefont
  {{Spergel}}}, \bibinfo {author} {\bibfnamefont {G.~S.}\ \bibnamefont
  {{Tucker}}}, \bibinfo {author} {\bibfnamefont {D.~T.}\ \bibnamefont
  {{Wilkinson}}}, \bibinfo {author} {\bibfnamefont {E.}~\bibnamefont
  {{Wollack}}}, \ and\ \bibinfo {author} {\bibfnamefont {E.~L.}\ \bibnamefont
  {{Wright}}},\ }\href {\doibase 10.1086/345346} {\bibfield  {journal}
  {\bibinfo  {journal} {Astrophys.\ J.}\ }\textbf {\bibinfo {volume} {583}},\
  \bibinfo {pages} {1} (\bibinfo {year} {2003})},\ \Eprint
  {http://arxiv.org/abs/arXiv:astro-ph/0301158} {arXiv:astro-ph/0301158}
  \BibitemShut {NoStop}%
\bibitem [{\citenamefont {{Tauber}}\ \emph {et~al.}(2010)\citenamefont
  {{Tauber}}, \citenamefont {{Mandolesi}}, \citenamefont {{Puget}} \emph
  {et~al.}}]{tauber:2010}%
  \BibitemOpen
  \bibfield  {author} {\bibinfo {author} {\bibfnamefont {J.~A.}\ \bibnamefont
  {{Tauber}}}, \bibinfo {author} {\bibfnamefont {N.}~\bibnamefont
  {{Mandolesi}}}, \bibinfo {author} {\bibfnamefont {J.}~\bibnamefont
  {{Puget}}},  \emph {et~al.},\ }\href {\doibase 10.1051/0004-6361/200912983}
  {\bibfield  {journal} {\bibinfo  {journal} {Astron.\ \& Astrophys.}\ }\textbf
  {\bibinfo {volume} {520}} (\bibinfo {year} {2010}),\
  10.1051/0004-6361/200912983}\BibitemShut {NoStop}%
\bibitem [{\citenamefont {Feeney}\ \emph
  {et~al.}(2011{\natexlab{a}})\citenamefont {Feeney}, \citenamefont {Johnson},
  \citenamefont {Mortlock},\ and\ \citenamefont {Peiris}}]{feeney:2011a}%
  \BibitemOpen
  \bibfield  {author} {\bibinfo {author} {\bibfnamefont {S.~M.}\ \bibnamefont
  {Feeney}}, \bibinfo {author} {\bibfnamefont {M.~C.}\ \bibnamefont {Johnson}},
  \bibinfo {author} {\bibfnamefont {D.~J.}\ \bibnamefont {Mortlock}}, \ and\
  \bibinfo {author} {\bibfnamefont {H.~V.}\ \bibnamefont {Peiris}},\ }\href
  {\doibase 10.1103/PhysRevD.84.043507} {\bibfield  {journal} {\bibinfo
  {journal} {Phys.\ Rev.\ D.}\ }\textbf {\bibinfo {volume} {D84}},\ \bibinfo
  {pages} {043507} (\bibinfo {year} {2011}{\natexlab{a}})},\ \Eprint
  {http://arxiv.org/abs/1012.3667} {arXiv:1012.3667 [astro-ph.CO]} \BibitemShut
  {NoStop}%
\bibitem [{\citenamefont {Feeney}\ \emph
  {et~al.}(2011{\natexlab{b}})\citenamefont {Feeney}, \citenamefont {Johnson},
  \citenamefont {Mortlock},\ and\ \citenamefont {Peiris}}]{feeney:2011b}%
  \BibitemOpen
  \bibfield  {author} {\bibinfo {author} {\bibfnamefont {S.~M.}\ \bibnamefont
  {Feeney}}, \bibinfo {author} {\bibfnamefont {M.~C.}\ \bibnamefont {Johnson}},
  \bibinfo {author} {\bibfnamefont {D.~J.}\ \bibnamefont {Mortlock}}, \ and\
  \bibinfo {author} {\bibfnamefont {H.~V.}\ \bibnamefont {Peiris}},\ }\href
  {\doibase 10.1103/PhysRevLett.107.071301} {\bibfield  {journal} {\bibinfo
  {journal} {Phys.\ Rev.\ Lett.}\ }\textbf {\bibinfo {volume} {107}},\ \bibinfo
  {pages} {071301} (\bibinfo {year} {2011}{\natexlab{b}})},\ \Eprint
  {http://arxiv.org/abs/1012.1995} {arXiv:1012.1995 [astro-ph.CO]} \BibitemShut
  {NoStop}%
\bibitem [{\citenamefont {{Marinucci}}\ \emph {et~al.}(2008)\citenamefont
  {{Marinucci}}, \citenamefont {{Pietrobon}}, \citenamefont {{Balbi}},
  \citenamefont {{Baldi}}, \citenamefont {{Cabella}}, \citenamefont
  {{Kerkyacharian}}, \citenamefont {{Natoli}}, \citenamefont {{Picard}},\ and\
  \citenamefont {{Vittorio}}}]{marinucci:2008}%
  \BibitemOpen
  \bibfield  {author} {\bibinfo {author} {\bibfnamefont {D.}~\bibnamefont
  {{Marinucci}}}, \bibinfo {author} {\bibfnamefont {D.}~\bibnamefont
  {{Pietrobon}}}, \bibinfo {author} {\bibfnamefont {A.}~\bibnamefont
  {{Balbi}}}, \bibinfo {author} {\bibfnamefont {P.}~\bibnamefont {{Baldi}}},
  \bibinfo {author} {\bibfnamefont {P.}~\bibnamefont {{Cabella}}}, \bibinfo
  {author} {\bibfnamefont {G.}~\bibnamefont {{Kerkyacharian}}}, \bibinfo
  {author} {\bibfnamefont {P.}~\bibnamefont {{Natoli}}}, \bibinfo {author}
  {\bibfnamefont {D.}~\bibnamefont {{Picard}}}, \ and\ \bibinfo {author}
  {\bibfnamefont {N.}~\bibnamefont {{Vittorio}}},\ }\href {\doibase
  10.1111/j.1365-2966.2007.12550.x} {\bibfield  {journal} {\bibinfo  {journal}
  {Mon.\ Not.\ Roy.\ Astron.\ Soc.}\ }\textbf {\bibinfo {volume} {383}},\
  \bibinfo {pages} {539} (\bibinfo {year} {2008})},\ \Eprint
  {http://arxiv.org/abs/arXiv:0707.0844} {arXiv:0707.0844} \BibitemShut
  {NoStop}%
\bibitem [{\citenamefont {Scodeller}\ \emph {et~al.}(2011)\citenamefont
  {Scodeller}, \citenamefont {Rudjord}, \citenamefont {Hansen}, \citenamefont
  {Marinucci}, \citenamefont {Geller} \emph {et~al.}}]{scodeller:2010}%
  \BibitemOpen
  \bibfield  {author} {\bibinfo {author} {\bibfnamefont {S.}~\bibnamefont
  {Scodeller}}, \bibinfo {author} {\bibfnamefont {O.}~\bibnamefont {Rudjord}},
  \bibinfo {author} {\bibfnamefont {F.}~\bibnamefont {Hansen}}, \bibinfo
  {author} {\bibfnamefont {D.}~\bibnamefont {Marinucci}}, \bibinfo {author}
  {\bibfnamefont {D.}~\bibnamefont {Geller}},  \emph {et~al.},\ }\href
  {\doibase 10.1088/0004-637X/733/2/121} {\bibfield  {journal} {\bibinfo
  {journal} {Astrophys.J.}\ }\textbf {\bibinfo {volume} {733}},\ \bibinfo
  {pages} {121} (\bibinfo {year} {2011})},\ \Eprint
  {http://arxiv.org/abs/1004.5576} {arXiv:1004.5576 [astro-ph.CO]} \BibitemShut
  {NoStop}%
\bibitem [{\citenamefont {Wiaux}\ \emph {et~al.}(2008)\citenamefont {Wiaux},
  \citenamefont {McEwen}, \citenamefont {Vandergheynst},\ and\ \citenamefont
  {Blanc}}]{wiaux:2007:sdw}%
  \BibitemOpen
  \bibfield  {author} {\bibinfo {author} {\bibfnamefont {Y.}~\bibnamefont
  {Wiaux}}, \bibinfo {author} {\bibfnamefont {J.~D.}\ \bibnamefont {McEwen}},
  \bibinfo {author} {\bibfnamefont {P.}~\bibnamefont {Vandergheynst}}, \ and\
  \bibinfo {author} {\bibfnamefont {O.}~\bibnamefont {Blanc}},\ }\href
  {\doibase 10.1111/j.1365-2966.2008.13448.x} {\bibfield  {journal} {\bibinfo
  {journal} {Mon.\ Not.\ Roy.\ Astron.\ Soc.}\ }\textbf {\bibinfo {volume}
  {388}},\ \bibinfo {pages} {770} (\bibinfo {year} {2008})},\ \Eprint
  {http://arxiv.org/abs/arXiv:0712.3519} {arXiv:0712.3519} \BibitemShut
  {NoStop}%
\bibitem [{\citenamefont {Tegmark}\ and\ \citenamefont
  {de~Oliveira-Costa}(1998)}]{tegmark:1998}%
  \BibitemOpen
  \bibfield  {author} {\bibinfo {author} {\bibfnamefont {M.}~\bibnamefont
  {Tegmark}}\ and\ \bibinfo {author} {\bibfnamefont {A.}~\bibnamefont
  {de~Oliveira-Costa}},\ }\href@noop {} {\bibfield  {journal} {\bibinfo
  {journal} {Astrophys.\ J.\ Lett.}\ }\textbf {\bibinfo {volume} {500}},\
  \bibinfo {pages} {L83} (\bibinfo {year} {1998})},\ \Eprint
  {http://arxiv.org/abs/astro-ph/9802123} {astro-ph/9802123} \BibitemShut
  {NoStop}%
\bibitem [{\citenamefont {Haehnelt}\ and\ \citenamefont
  {Tegmark}(1996)}]{haehnelt:1995}%
  \BibitemOpen
  \bibfield  {author} {\bibinfo {author} {\bibfnamefont {M.~G.}\ \bibnamefont
  {Haehnelt}}\ and\ \bibinfo {author} {\bibfnamefont {M.}~\bibnamefont
  {Tegmark}},\ }\href@noop {} {\bibfield  {journal} {\bibinfo  {journal} {Mon.\
  Not.\ Roy.\ Astron.\ Soc.}\ }\textbf {\bibinfo {volume} {279}},\ \bibinfo
  {pages} {545} (\bibinfo {year} {1996})},\ \Eprint
  {http://arxiv.org/abs/astro-ph/9507077} {astro-ph/9507077} \BibitemShut
  {NoStop}%
\bibitem [{\citenamefont {Sanz}\ \emph {et~al.}(2001)\citenamefont {Sanz},
  \citenamefont {Herranz},\ and\ \citenamefont
  {Mart\'{\i}nez-Gonz\'{a}lez}}]{sanz:2001}%
  \BibitemOpen
  \bibfield  {author} {\bibinfo {author} {\bibfnamefont {J.~L.}\ \bibnamefont
  {Sanz}}, \bibinfo {author} {\bibfnamefont {D.}~\bibnamefont {Herranz}}, \
  and\ \bibinfo {author} {\bibfnamefont {E.}~\bibnamefont
  {Mart\'{\i}nez-Gonz\'{a}lez}},\ }\href@noop {} {\bibfield  {journal}
  {\bibinfo  {journal} {Astrophys.\ J.}\ }\textbf {\bibinfo {volume} {552}},\
  \bibinfo {pages} {484} (\bibinfo {year} {2001})},\ \Eprint
  {http://arxiv.org/abs/astro-ph/0107384} {astro-ph/0107384} \BibitemShut
  {NoStop}%
\bibitem [{\citenamefont {Herranz}\ \emph {et~al.}(2002)\citenamefont
  {Herranz}, \citenamefont {Sanz}, \citenamefont {Barreiro},\ and\
  \citenamefont {Mart\'{\i}nez-Gonz\'{a}lez}}]{herranz:2002}%
  \BibitemOpen
  \bibfield  {author} {\bibinfo {author} {\bibfnamefont {D.}~\bibnamefont
  {Herranz}}, \bibinfo {author} {\bibfnamefont {J.~L.}\ \bibnamefont {Sanz}},
  \bibinfo {author} {\bibfnamefont {R.~B.}\ \bibnamefont {Barreiro}}, \ and\
  \bibinfo {author} {\bibfnamefont {E.}~\bibnamefont
  {Mart\'{\i}nez-Gonz\'{a}lez}},\ }\href@noop {} {\bibfield  {journal}
  {\bibinfo  {journal} {Astrophys.\ J.}\ }\textbf {\bibinfo {volume} {580}},\
  \bibinfo {pages} {610} (\bibinfo {year} {2002})},\ \Eprint
  {http://arxiv.org/abs/astro-ph/0204149} {astro-ph/0204149} \BibitemShut
  {NoStop}%
\bibitem [{\citenamefont {{Barreiro}}\ \emph {et~al.}(2003)\citenamefont
  {{Barreiro}}, \citenamefont {{Sanz}}, \citenamefont {{Herranz}},\ and\
  \citenamefont {{Mart{\'{\i}}nez-Gonz{\'a}lez}}}]{barreiro:2003}%
  \BibitemOpen
  \bibfield  {author} {\bibinfo {author} {\bibfnamefont {R.~B.}\ \bibnamefont
  {{Barreiro}}}, \bibinfo {author} {\bibfnamefont {J.~L.}\ \bibnamefont
  {{Sanz}}}, \bibinfo {author} {\bibfnamefont {D.}~\bibnamefont {{Herranz}}}, \
  and\ \bibinfo {author} {\bibfnamefont {E.}~\bibnamefont
  {{Mart{\'{\i}}nez-Gonz{\'a}lez}}},\ }\href {\doibase
  10.1046/j.1365-8711.2003.06520.x} {\bibfield  {journal} {\bibinfo  {journal}
  {Mon.\ Not.\ Roy.\ Astron.\ Soc.}\ }\textbf {\bibinfo {volume} {342}},\
  \bibinfo {pages} {119} (\bibinfo {year} {2003})},\ \Eprint
  {http://arxiv.org/abs/astro-ph/0302245} {astro-ph/0302245} \BibitemShut
  {NoStop}%
\bibitem [{\citenamefont {Schaefer}\ \emph {et~al.}(2006)\citenamefont
  {Schaefer}, \citenamefont {Pfrommer}, \citenamefont {Hell},\ and\
  \citenamefont {Bartelmann}}]{schaefer:2004}%
  \BibitemOpen
  \bibfield  {author} {\bibinfo {author} {\bibfnamefont {B.~M.}\ \bibnamefont
  {Schaefer}}, \bibinfo {author} {\bibfnamefont {C.}~\bibnamefont {Pfrommer}},
  \bibinfo {author} {\bibfnamefont {R.~M.}\ \bibnamefont {Hell}}, \ and\
  \bibinfo {author} {\bibfnamefont {M.}~\bibnamefont {Bartelmann}},\ }\href
  {\doibase 10.1111/j.1365-2966.2006.10622.x} {\bibfield  {journal} {\bibinfo
  {journal} {Mon.\ Not.\ Roy.\ Astron.\ Soc.}\ }\textbf {\bibinfo {volume}
  {370}},\ \bibinfo {pages} {1713} (\bibinfo {year} {2006})},\ \Eprint
  {http://arxiv.org/abs/astro-ph/0407090} {astro-ph/0407090} \BibitemShut
  {NoStop}%
\bibitem [{\citenamefont {Malte~Schafer}\ and\ \citenamefont
  {Bartelmann}(2007)}]{schaefer:2006}%
  \BibitemOpen
  \bibfield  {author} {\bibinfo {author} {\bibfnamefont {B.}~\bibnamefont
  {Malte~Schafer}}\ and\ \bibinfo {author} {\bibfnamefont {M.}~\bibnamefont
  {Bartelmann}},\ }\href {\doibase 10.1111/j.1365-2966.2007.11596.x} {\bibfield
   {journal} {\bibinfo  {journal} {Mon.Not.Roy.Astron.Soc.}\ }\textbf {\bibinfo
  {volume} {377}},\ \bibinfo {pages} {253} (\bibinfo {year} {2007})},\ \Eprint
  {http://arxiv.org/abs/astro-ph/0602406} {arXiv:astro-ph/0602406 [astro-ph]}
  \BibitemShut {NoStop}%
\bibitem [{\citenamefont {McEwen}\ \emph {et~al.}(2008)\citenamefont {McEwen},
  \citenamefont {Hobson},\ and\ \citenamefont {Lasenby}}]{mcewen:2006:filters}%
  \BibitemOpen
  \bibfield  {author} {\bibinfo {author} {\bibfnamefont {J.~D.}\ \bibnamefont
  {McEwen}}, \bibinfo {author} {\bibfnamefont {M.~P.}\ \bibnamefont {Hobson}},
  \ and\ \bibinfo {author} {\bibfnamefont {A.~N.}\ \bibnamefont {Lasenby}},\
  }\href {\doibase 10.1109/TSP.2008.923198} {\bibfield  {journal} {\bibinfo
  {journal} {IEEE Trans.\ Sig.\ Proc.}\ }\textbf {\bibinfo {volume} {56}},\
  \bibinfo {pages} {3813} (\bibinfo {year} {2008})},\ \Eprint
  {http://arxiv.org/abs/astro-ph/0612688} {astro-ph/0612688} \BibitemShut
  {NoStop}%
\bibitem [{\citenamefont {Driscoll}\ and\ \citenamefont
  {Healy}(1994)}]{driscoll:1994}%
  \BibitemOpen
  \bibfield  {author} {\bibinfo {author} {\bibfnamefont {J.~R.}\ \bibnamefont
  {Driscoll}}\ and\ \bibinfo {author} {\bibfnamefont {D.~M.~J.}\ \bibnamefont
  {Healy}},\ }\href@noop {} {\bibfield  {journal} {\bibinfo  {journal}
  {Advances in Applied Mathematics}\ }\textbf {\bibinfo {volume} {15}},\
  \bibinfo {pages} {202} (\bibinfo {year} {1994})}\BibitemShut {NoStop}%
\bibitem [{\citenamefont {G\'{o}rski}\ \emph {et~al.}(2005)\citenamefont
  {G\'{o}rski}, \citenamefont {Hivon}, \citenamefont {Banday}, \citenamefont
  {Wandelt}, \citenamefont {Hansen}, \citenamefont {Reinecke},\ and\
  \citenamefont {Bartelmann}}]{gorski:2005}%
  \BibitemOpen
  \bibfield  {author} {\bibinfo {author} {\bibfnamefont {K.~M.}\ \bibnamefont
  {G\'{o}rski}}, \bibinfo {author} {\bibfnamefont {E.}~\bibnamefont {Hivon}},
  \bibinfo {author} {\bibfnamefont {A.~J.}\ \bibnamefont {Banday}}, \bibinfo
  {author} {\bibfnamefont {B.~D.}\ \bibnamefont {Wandelt}}, \bibinfo {author}
  {\bibfnamefont {F.~K.}\ \bibnamefont {Hansen}}, \bibinfo {author}
  {\bibfnamefont {M.}~\bibnamefont {Reinecke}}, \ and\ \bibinfo {author}
  {\bibfnamefont {M.}~\bibnamefont {Bartelmann}},\ }\href@noop {} {\bibfield
  {journal} {\bibinfo  {journal} {Astrophys.\ J.}\ }\textbf {\bibinfo {volume}
  {622}},\ \bibinfo {pages} {759} (\bibinfo {year} {2005})},\ \Eprint
  {http://arxiv.org/abs/astro-ph/0409513} {astro-ph/0409513} \BibitemShut
  {NoStop}%
\bibitem [{\citenamefont {Doroshkevich}\ \emph {et~al.}(2005)\citenamefont
  {Doroshkevich}, \citenamefont {Naselsky}, \citenamefont {Verkhodanov},
  \citenamefont {Novikov}, \citenamefont {Turchaninov}, \citenamefont
  {Novikov}, \citenamefont {Christensen},\ and\ \citenamefont
  {Chiang}}]{doroshkevich:2005}%
  \BibitemOpen
  \bibfield  {author} {\bibinfo {author} {\bibfnamefont {A.~G.}\ \bibnamefont
  {Doroshkevich}}, \bibinfo {author} {\bibfnamefont {P.~D.}\ \bibnamefont
  {Naselsky}}, \bibinfo {author} {\bibfnamefont {O.~V.}\ \bibnamefont
  {Verkhodanov}}, \bibinfo {author} {\bibfnamefont {D.~I.}\ \bibnamefont
  {Novikov}}, \bibinfo {author} {\bibfnamefont {V.~I.}\ \bibnamefont
  {Turchaninov}}, \bibinfo {author} {\bibfnamefont {I.~D.}\ \bibnamefont
  {Novikov}}, \bibinfo {author} {\bibfnamefont {P.~R.}\ \bibnamefont
  {Christensen}}, \ and\ \bibinfo {author} {\bibfnamefont {L.~Y.}\ \bibnamefont
  {Chiang}},\ }\href@noop {} {\bibfield  {journal} {\bibinfo  {journal} {Int.
  J. Mod. Phys. D.}\ }\textbf {\bibinfo {volume} {14}},\ \bibinfo {pages} {275}
  (\bibinfo {year} {2005})},\ \Eprint {http://arxiv.org/abs/astro-ph/0305537}
  {astro-ph/0305537} \BibitemShut {NoStop}%
\bibitem [{\citenamefont {McEwen}\ and\ \citenamefont
  {Wiaux}(2011)}]{mcewen:fssht}%
  \BibitemOpen
  \bibfield  {author} {\bibinfo {author} {\bibfnamefont {J.~D.}\ \bibnamefont
  {McEwen}}\ and\ \bibinfo {author} {\bibfnamefont {Y.}~\bibnamefont {Wiaux}},\
  }\href {\doibase 10.1109/TSP.2011.2166394} {\bibfield  {journal} {\bibinfo
  {journal} {IEEE Trans.\ Sig.\ Proc.}\ }\textbf {\bibinfo {volume} {59}},\
  \bibinfo {pages} {5876} (\bibinfo {year} {2011})},\ \Eprint
  {http://arxiv.org/abs/arXiv:1110.6298} {arXiv:1110.6298} \BibitemShut
  {NoStop}%
\bibitem [{\citenamefont {Risbo}(1996)}]{risbo:1996}%
  \BibitemOpen
  \bibfield  {author} {\bibinfo {author} {\bibfnamefont {T.}~\bibnamefont
  {Risbo}},\ }\href@noop {} {\bibfield  {journal} {\bibinfo  {journal} {J.\
  Geodesy}\ }\textbf {\bibinfo {volume} {70}},\ \bibinfo {pages} {383}
  (\bibinfo {year} {1996})}\BibitemShut {NoStop}%
\bibitem [{\citenamefont {Wandelt}\ and\ \citenamefont
  {G\'{o}rski}(2001)}]{wandelt:2001}%
  \BibitemOpen
  \bibfield  {author} {\bibinfo {author} {\bibfnamefont {B.~D.}\ \bibnamefont
  {Wandelt}}\ and\ \bibinfo {author} {\bibfnamefont {K.~M.}\ \bibnamefont
  {G\'{o}rski}},\ }\href@noop {} {\bibfield  {journal} {\bibinfo  {journal}
  {Phys.\ Rev.\ D.}\ }\textbf {\bibinfo {volume} {63}},\ \bibinfo {pages}
  {123002} (\bibinfo {year} {2001})},\ \Eprint
  {http://arxiv.org/abs/astro-ph/0008227} {astro-ph/0008227} \BibitemShut
  {NoStop}%
\bibitem [{\citenamefont {McEwen}\ \emph {et~al.}(2007)\citenamefont {McEwen},
  \citenamefont {Hobson}, \citenamefont {Mortlock},\ and\ \citenamefont
  {Lasenby}}]{mcewen:2006:fcswt}%
  \BibitemOpen
  \bibfield  {author} {\bibinfo {author} {\bibfnamefont {J.~D.}\ \bibnamefont
  {McEwen}}, \bibinfo {author} {\bibfnamefont {M.~P.}\ \bibnamefont {Hobson}},
  \bibinfo {author} {\bibfnamefont {D.~J.}\ \bibnamefont {Mortlock}}, \ and\
  \bibinfo {author} {\bibfnamefont {A.~N.}\ \bibnamefont {Lasenby}},\ }\href
  {\doibase 10.1109/TSP.2006.887148} {\bibfield  {journal} {\bibinfo  {journal}
  {IEEE Trans.\ Sig.\ Proc.}\ }\textbf {\bibinfo {volume} {55}},\ \bibinfo
  {pages} {520} (\bibinfo {year} {2007})},\ \Eprint
  {http://arxiv.org/abs/astro-ph/0506308} {astro-ph/0506308} \BibitemShut
  {NoStop}%
\bibitem [{\citenamefont {{Jarosik}}\ \emph {et~al.}(2011)\citenamefont
  {{Jarosik}}, \citenamefont {{Bennett}}, \citenamefont {{Dunkley}},
  \citenamefont {{Gold}}, \citenamefont {{Greason}}, \citenamefont {{Halpern}},
  \citenamefont {{Hill}}, \citenamefont {{Hinshaw}}, \citenamefont {{Kogut}},
  \citenamefont {{Komatsu}}, \citenamefont {{Larson}}, \citenamefont {{Limon}},
  \citenamefont {{Meyer}}, \citenamefont {{Nolta}}, \citenamefont {{Odegard}},
  \citenamefont {{Page}}, \citenamefont {{Smith}}, \citenamefont {{Spergel}},
  \citenamefont {{Tucker}}, \citenamefont {{Weiland}}, \citenamefont
  {{Wollack}},\ and\ \citenamefont {{Wright}}}]{jaroski:2010}%
  \BibitemOpen
  \bibfield  {author} {\bibinfo {author} {\bibfnamefont {N.}~\bibnamefont
  {{Jarosik}}}, \bibinfo {author} {\bibfnamefont {C.~L.}\ \bibnamefont
  {{Bennett}}}, \bibinfo {author} {\bibfnamefont {J.}~\bibnamefont
  {{Dunkley}}}, \bibinfo {author} {\bibfnamefont {B.}~\bibnamefont {{Gold}}},
  \bibinfo {author} {\bibfnamefont {M.~R.}\ \bibnamefont {{Greason}}}, \bibinfo
  {author} {\bibfnamefont {M.}~\bibnamefont {{Halpern}}}, \bibinfo {author}
  {\bibfnamefont {R.~S.}\ \bibnamefont {{Hill}}}, \bibinfo {author}
  {\bibfnamefont {G.}~\bibnamefont {{Hinshaw}}}, \bibinfo {author}
  {\bibfnamefont {A.}~\bibnamefont {{Kogut}}}, \bibinfo {author} {\bibfnamefont
  {E.}~\bibnamefont {{Komatsu}}}, \bibinfo {author} {\bibfnamefont
  {D.}~\bibnamefont {{Larson}}}, \bibinfo {author} {\bibfnamefont
  {M.}~\bibnamefont {{Limon}}}, \bibinfo {author} {\bibfnamefont {S.~S.}\
  \bibnamefont {{Meyer}}}, \bibinfo {author} {\bibfnamefont {M.~R.}\
  \bibnamefont {{Nolta}}}, \bibinfo {author} {\bibfnamefont {N.}~\bibnamefont
  {{Odegard}}}, \bibinfo {author} {\bibfnamefont {L.}~\bibnamefont {{Page}}},
  \bibinfo {author} {\bibfnamefont {K.~M.}\ \bibnamefont {{Smith}}}, \bibinfo
  {author} {\bibfnamefont {D.~N.}\ \bibnamefont {{Spergel}}}, \bibinfo {author}
  {\bibfnamefont {G.~S.}\ \bibnamefont {{Tucker}}}, \bibinfo {author}
  {\bibfnamefont {J.~L.}\ \bibnamefont {{Weiland}}}, \bibinfo {author}
  {\bibfnamefont {E.}~\bibnamefont {{Wollack}}}, \ and\ \bibinfo {author}
  {\bibfnamefont {E.~L.}\ \bibnamefont {{Wright}}},\ }\href {\doibase
  10.1088/0067-0049/192/2/14} {\bibfield  {journal} {\bibinfo  {journal}
  {Astrophys.\ J.\ Supp.}\ }\textbf {\bibinfo {volume} {192}},\ \bibinfo {eid}
  {14} (\bibinfo {year} {2011})},\ \Eprint {http://arxiv.org/abs/1001.4744}
  {arXiv:1001.4744 [astro-ph.CO]} \BibitemShut {NoStop}%
\bibitem [{\citenamefont {{Gold}}\ \emph {et~al.}(2011)\citenamefont {{Gold}},
  \citenamefont {{Odegard}}, \citenamefont {{Weiland}}, \citenamefont {{Hill}},
  \citenamefont {{Kogut}}, \citenamefont {{Bennett}}, \citenamefont
  {{Hinshaw}}, \citenamefont {{Chen}}, \citenamefont {{Dunkley}}, \citenamefont
  {{Halpern}}, \citenamefont {{Jarosik}}, \citenamefont {{Komatsu}},
  \citenamefont {{Larson}}, \citenamefont {{Limon}}, \citenamefont {{Meyer}},
  \citenamefont {{Nolta}}, \citenamefont {{Page}}, \citenamefont {{Smith}},
  \citenamefont {{Spergel}}, \citenamefont {{Tucker}}, \citenamefont
  {{Wollack}},\ and\ \citenamefont {{Wright}}}]{gold:2011}%
  \BibitemOpen
  \bibfield  {author} {\bibinfo {author} {\bibfnamefont {B.}~\bibnamefont
  {{Gold}}}, \bibinfo {author} {\bibfnamefont {N.}~\bibnamefont {{Odegard}}},
  \bibinfo {author} {\bibfnamefont {J.~L.}\ \bibnamefont {{Weiland}}}, \bibinfo
  {author} {\bibfnamefont {R.~S.}\ \bibnamefont {{Hill}}}, \bibinfo {author}
  {\bibfnamefont {A.}~\bibnamefont {{Kogut}}}, \bibinfo {author} {\bibfnamefont
  {C.~L.}\ \bibnamefont {{Bennett}}}, \bibinfo {author} {\bibfnamefont
  {G.}~\bibnamefont {{Hinshaw}}}, \bibinfo {author} {\bibfnamefont
  {X.}~\bibnamefont {{Chen}}}, \bibinfo {author} {\bibfnamefont
  {J.}~\bibnamefont {{Dunkley}}}, \bibinfo {author} {\bibfnamefont
  {M.}~\bibnamefont {{Halpern}}}, \bibinfo {author} {\bibfnamefont
  {N.}~\bibnamefont {{Jarosik}}}, \bibinfo {author} {\bibfnamefont
  {E.}~\bibnamefont {{Komatsu}}}, \bibinfo {author} {\bibfnamefont
  {D.}~\bibnamefont {{Larson}}}, \bibinfo {author} {\bibfnamefont
  {M.}~\bibnamefont {{Limon}}}, \bibinfo {author} {\bibfnamefont {S.~S.}\
  \bibnamefont {{Meyer}}}, \bibinfo {author} {\bibfnamefont {M.~R.}\
  \bibnamefont {{Nolta}}}, \bibinfo {author} {\bibfnamefont {L.}~\bibnamefont
  {{Page}}}, \bibinfo {author} {\bibfnamefont {K.~M.}\ \bibnamefont {{Smith}}},
  \bibinfo {author} {\bibfnamefont {D.~N.}\ \bibnamefont {{Spergel}}}, \bibinfo
  {author} {\bibfnamefont {G.~S.}\ \bibnamefont {{Tucker}}}, \bibinfo {author}
  {\bibfnamefont {E.}~\bibnamefont {{Wollack}}}, \ and\ \bibinfo {author}
  {\bibfnamefont {E.~L.}\ \bibnamefont {{Wright}}},\ }\href {\doibase
  10.1088/0067-0049/192/2/15} {\bibfield  {journal} {\bibinfo  {journal}
  {Astrophys.\ J.\ Supp.}\ }\textbf {\bibinfo {volume} {192}},\ \bibinfo {eid}
  {15} (\bibinfo {year} {2011})},\ \Eprint {http://arxiv.org/abs/1001.4555}
  {arXiv:1001.4555 [astro-ph.GA]} \BibitemShut {NoStop}%
\end{thebibliography}%

\end{document}